\documentclass[
  journal=largetwo,
  manuscript=article-type,
  year=2020,
  volume=37,
]{cup-journal}

\usepackage{amsmath}
\usepackage{gensymb}
\usepackage[nopatch]{microtype}
\usepackage{booktabs}
\usepackage[utf8]{inputenc}
\usepackage{rotating}
\usepackage{adjustbox}
\usepackage{afterpage}
\usepackage{subcaption}
\usepackage{stfloats}
\usepackage{xcolor}
\usepackage{lineno}
\usepackage{placeins}

\newcommand{\hi}{H{\sc i}}
\newcommand{\hii}{H{\sc ii}}

\newcommand{\msun}{$M_{\odot}$}
\newcommand{\mstar}{$M_{\star}$}
\newcommand{\mhi}{$M$\textsubscript{\sc{HI}}}
\newcommand{\shi}{$S$\textsubscript{\sc{HI}}}
\newcommand{\rhi}{$R$\textsubscript{\sc{HI}}}
\newcommand{\kms}{km s$^{-1}$}
\newcommand{\yr}{yr$^{-1}$}
\newcommand{\tenpow}[1]{$\times 10^{#1}$}
\newcommand{\PM}{$\pm$}
\newcommand{\Aflux}{$A$\textsubscript{flux}}
\newcommand{\Aspec}{$A$\textsubscript{spec}}
\newcommand{\AoD}{$A$\textsubscript{1D}}
\newcommand{\AtD}{$A$\textsubscript{3D}}
\newcommand{\red}[1]{\textcolor{black}{#1}}

\title{The Distribution of Atomic Hydrogen in the Host Galaxies of FRBs}

\author{H. Roxburgh}
\affiliation{International Centre for Radio Astronomy Research, Curtin University, Bentley, WA 6102, Australia}
\email[H. Roxburgh]{hugh.roxburgh@postgrad.curtin.edu.au}

\author{M. Glowacki}
\affiliation{International Centre for Radio Astronomy Research, Curtin University, Bentley, WA 6102, Australia}
\alsoaffiliation{Institute for Astronomy, University of Edinburgh, Royal Observatory, Edinburgh, EH9 3HJ, United Kingdom}
\alsoaffiliation{Inter-University Institute for Data Intensive Astronomy, Department of Astronomy, University of Cape Town, Cape Town, South Africa}

\author{A. Bera}
\affiliation{International Centre for Radio Astronomy Research, Curtin University, Bentley, WA 6102, Australia}

\author{C. W. James}
\affiliation{International Centre for Radio Astronomy Research, Curtin University, Bentley, WA 6102, Australia}

\author{N. Deg}
\affiliation{Department of Physics, Engineering Physics, and Astronomy,Queen's University, Kingston ON K7L 3N6, Canada}

\author{Q. Huang}
\affiliation{Kavli Institute for Astronomy and Astrophysics, Peking University, Beijing 100871, China; Department of Astronomy, School of Physics, Peking University, Beijing 100871, China}

\author{K. Lee-Waddell}
\affiliation{Australian SKA Regional Centre (AusSRC) - The University of Western Australia, 35 Stirling Highway, Crawley WA 6009, Australia}
\alsoaffiliation{CSIRO Space and Astronomy, PO Box 1130, Bentley WA 6102, Australia}
\alsoaffiliation{International Centre for Radio Astronomy Research, Curtin University, Bentley, WA 6102, Australia}

\author{J. Wang}
\affiliation{Kavli Institute for Astronomy and Astrophysics, Peking University, Beijing 100871, China; Department of Astronomy, School of Physics, Peking University, Beijing 100871, China}

\author{M. Caleb}
\affiliation{Sydney Institute for Astronomy, School of Physics, The University of Sydney, NSW 2006, Australia}

\author{A.~T.~Deller}
\affiliation{Centre for Astrophysics and Supercomputing, Swinburne University of Technology, Hawthorn, VIC, 3122, Australia}

\author{L. N. Driessen}
\affiliation{Sydney Institute for Astronomy, School of Physics, The University of Sydney, NSW 2006, Australia}

\author{A. C. Gordon}
\affiliation{Center for Interdisciplinary Exploration and Research in Astrophysics (CIERA) and Department of Physics and Astronomy, Northwestern University, Evanston, IL 60208, USA}

\author{K.~M.~Hess}
\affiliation{Department of Space, Earth and Environment, Chalmers University of Technology, Onsala Space Observatory, SE-43992 Onsala, Sweden}
\alsoaffiliation{ASTRON, The Netherlands Institute for Radio Astronomy, Postbus 2, 7990 AA, Dwingeloo, The Netherlands}

\author{J.~X.~Prochaska}
\affiliation{Department of Astronomy and Astrophysics, University of California, Santa Cruz, CA 95064, USA}
\alsoaffiliation{Kavli Institute for the Physics and Mathematics of the Universe (Kavli IPMU), 5-1-5 Kashiwanoha, Kashiwa, 277-8583, Japan}
\alsoaffiliation{Division of Science, National Astronomical Observatory of Japan, 2-21-1 Osawa, Mitaka, Tokyo, 181-8588, Japan}

\author{R.~M.~Shannon}
\affiliation{Centre for Astrophysics and Supercomputing, Swinburne University of Technology, Hawthorn, VIC, 3122, Australia}

\author{Y. Wang}
\affiliation{Centre for Astrophysics and Supercomputing, Swinburne University of Technology, Hawthorn, VIC, 3122, Australia}

\author{Z. Wang}
\affiliation{International Centre for Radio Astronomy Research, Curtin University, Bentley, WA 6102, Australia}

\author{D. Yang}
\affiliation{Kavli Institute for Astronomy and Astrophysics, Peking University, Beijing 100871, China; Department of Astronomy, School of Physics, Peking University, Beijing 100871, China}

\begin{document}
% \linenumbers

\begin{abstract}
We probe the atomic hydrogen (\hi) emission from the host galaxies of fast radio bursts (FRBs) to investigate the emerging trend of disturbance and asymmetry in the population. Quadrupling the sample size, we detect \red{16 out of 17} new hosts in \hi, with the single non-detection arising in a galaxy known to be transitioning towards quiescence. With respect to typical local Universe galaxies, FRB hosts are generally massive in \hi\ (\mhi\ $>10^9$ \msun), which aligns with previous studies reporting that FRB hosts also tend to have high stellar masses and are star-forming. However, they span a broad range of other \hi\ derived properties. Using visual inspection alongside various asymmetry metrics, we identify \red{six} unambiguously settled host galaxies, demonstrating for the first time that a disturbed \hi\ morphology is not a universal feature of FRB host galaxies. However, we find another six that show clear signs of disturbance, \red{one borderline case,} and three which require deeper or more targeted observations to reach a conclusion; this brings the confirmed ratio of disturbed-to-settled FRB hosts to \red{11:6}. Given that roughly a 1:1 ratio is expected for random background galaxies of similar type, our observed ratio yields a p-value of \red{0.222. Therefore, we conclude that contrary to earlier indications, there is no statistically significant excess of \hi\ disturbance in this sample of FRB host galaxies with respect to the general galaxy population, and hence we find no evidence for a fundamental connection between FRB progenitor formation and merger-induced star formation activity.}

% Unlike earlier indications based on smaller samples, this no longer crosses the conventional threshold for statistical significance, though is still near enough to hint at a legitimate excess of disturbance among FRB hosts. Thus, an even larger sample size of FRB hosts observed in \hi\ is required to fully clarify whether the trend is genuine or still a consequence of low-number statistics --- a sample that upcoming data releases are well positioned to provide.
\end{abstract}

\section{Introduction}\label{sec:intro}

Fast radio bursts (FRBs) are energetic pulses of radio emission that generally last for a few milliseconds in timescale \citep{Lorimer2007}. Whilst their extragalactic origins are well established \citep{Chatterjee2017,Bannister2019}, their generation mechanisms remain unclear, despite significant advances in the field. Dozens of progenitor models have been proposed to explain the phenomena, from outbursting neutron stars to interacting black holes \citep[for reviews, see][]{Platts2019,Petroff2019}. The former, specifically young, highly magnetised neutron stars or magnetars, are favoured within the community due to observations of energetic FRBs found to be embedded within persistent radio sources (PRSs) \citep[e.g.][]{Metzger2017}. This link was strengthened in 2020 with the detection of an FRB from a magnetar within our galaxy \citep{Andersen2020,Bochenek2020}, although its isotropic-equivalent energy falls nearly 30 times lower than the next weakest FRB.

Since their discovery, the population of detected FRBs has rapidly expanded, primarily due to dedicated observations from the Australian Square Kilometre Array Pathfinder \citep[ASKAP;][]{ASKAP,Bannister2019,Shannon2024}, the Canadian Hydrogen Intensity Mapping Experiment \citep[CHIME;][]{CHIME,ChimeC1}, MeerKAT \citep{MeerKAT,MeerTRAP}, and the Deep Synoptic Array \citep[DSA;][]{Kocz2019,DSA,Connor2025}. Crucially, improvements to spatial resolution have allowed many FRBs to be localised to arcsecond or better accuracy, allowing for deep follow up photometry and spectroscopy of their host galaxies and environments \citep[e.g.][]{Bannister2019,Heintz2020,Mannings2021,Bhandari2022,Gordon2023,Ryder2023,Sharma2024,Amiri2025}. Such localisations are key to revealing the environments of FRBs and unlocking their potential for cosmological studies \citep[e.g.][]{McQuinn2014,Macquart2020,James2022,Connor2025}.

As of September 2025, over 100 host galaxies have been identified, spanning redshifts out to $z=$2.15. These galaxies tend to be star-forming spiral galaxies, though most other galactic parameters such as stellar mass and mass-weighted age are highly varied \citep{Bhandari2022,Gordon2023}. These observations broadly support a magnetar progenitor hypothesis, as magnetars are expected to trace star formation through a core-collapse supernovae generation channel \citep{Metzger2017}. However, a number of counterexamples, such as a localisation to a globular cluster \citep{Kirsten2022}, and several to quiescent galaxies \citep{Eftekhari2025,Shah2025}, muddy the waters. Furthermore, no statistically significant trend differing the hosts of repeating and apparently non-repeating FRBs has been observed \citep{Gordon2023,Sharma2024}, despite significant evidence suggesting fundamental differences in their respective origins \citep{ChimeC1}.

% It thus remains unclear how much direct value such population studies can lend to drawing a definitive conclusion on the progenitor problem.

Beyond analysis of the optical output of these host galaxies, a number of studies have been conducted on their atomic hydrogen (\hi) gas emission \citep{Michalowski21,Kaur2022,Glowacki_wallaby,Lee-waddell23,Glowacki_hi_frb_whats_your_z}. At redshifts greater than 0.1, the \hi\ line becomes weak and contaminated by RFI; with much of the localised FRB population lying beyond this redshift, only six host galaxies (excluding the Milky Way) have thus far been probed. All six of these galaxies have been detected in \hi, an expected result considering the star-forming nature of most hosts as derived from optical observations. Interestingly, a trend has emerged, with five of six of these detections exhibiting either a strongly asymmetric \hi\ spectrum or a highly disturbed spatial gas distribution. 
The single apparently symmetric profile, that of FRB 20211127I, was analysed with a low resolution and low signal-to-noise ratio (SNR) ASKAP observation, and thus its lack of disturbance was considered inconclusive \citep{Glowacki_wallaby}. 

Such features, which are well documented in the literature \citep[e.g.][]{Holwerda2011,Espada2011,Bok2019,Reynolds2020}, are attributed to recent or ongoing disruptions to the galactic environment. The gaseous disk is sensitive to both external and internal perturbative mechanisms, including tidal interactions and mergers, ram pressure stripping, and feedback from supernovae or active galactic nuclei \citep[AGN;][]{Yu2022}. Disturbed \hi\ distributions are not uncommon; many studies have found that a significant fraction of galaxies exhibit some sign of disturbance. Historical studies using single dish instruments focused on the asymmetry of the 1D profile; early estimates found that at least 50\% of galaxies exhibit spectra with some level of asymmetry \citep{Richter94,Haynes1998,Matthews1998}. However, more recent studies use more stringent cutoffs; including these results \citep{Espada2011,Watts2020,Reynolds2020,Yu2022} and reevaluating the historical ones find anywhere between 10-40\% of galaxies exhibit asymmetric \hi\ line profiles.

In comparison, few studies have investigated similar properties in the spatial distribution of resolved \hi\ galaxies \citep{Holwerda2011,vanEymeren}. \citet{Reynolds2020LVHIS} recently probed $\sim$120 galaxies from the LVHIS \citep{LVHIS}, VIVA \citep{VIVA}, and HALOGAS \citep{HALOGAS} surveys, finding similar rates of asymmetry. However, with a number of more advanced interferometric observatories currently in the process of conducting large \hi\ surveys of resolved galaxies --- such as the Widefield ASKAP L-band Legacy All-sky Blind surveY \citep[WALLABY,][]{WALLABY} on ASKAP, the \hi\ project within the MeerKAT International GigaHertz
Tiered Extragalactic Exploration survey \citep[MIGHTEE-HI,][]{MIGHTEEHI}\red{, and the Apertif HI surveys \citep{ApertifHI} on the Westerbork Synthesis Radio Telescope \citep[WRST,][]{WRSTapertif}} --- many more observations will be made to properly pin down morphological asymmetry in \hi\ disks. 

Overall, the exact fraction of disturbed galaxies measured by each study is highly dependent on the samples used (e.g., isolated vs galaxy pairs), \red{the observational properties (e.g. resolution and sensitivity)}, the method of quantification, and the point of cutoff in these methods above which a profile is considered disrupted. However, it can generally be said that disturbances are common, and are likely present in up to half of the galaxy population.

Thus, given the low number of hosts probed, the early trend towards asymmetry in the population --- with a ratio of 5:0 or 5:1 --- is not necessarily significant at this stage. However, if this trend perseveres as more host galaxies are observed in \hi, it may indicate a connection between FRBs and a recent enhancement of star formation due to galactic interaction \citep{Michalowski21}. 
Indeed, studies over the past few decades have identified higher star formation rates in merging galaxies \citep[e.g.][]{Alonso2004,Darg2010}. Such a scenario would lend further credence to so called `fast track' FRB progenitors such as magnetars, which are expected to trace star formation.  

To this end, we set forth to expand the sample size of FRB host galaxies observed in \hi, and search for signs of asymmetry in their spectral and spatial profiles. This paper is structured as follows: in Section \ref{sec:data}, we outline the sample of FRB host galaxy targets, and detail the observations and processing methods used to acquire \hi\ information. In Section \ref{sec:results}, we present our detections and the \hi\ properties of the sample. Section \ref{sec:asym} highlights the various methods we use to assess the disturbance/asymmetry of a detection. We evaluate each target in Section \ref{sec:individuals}, and discuss the implications of these findings in Section \ref{sec:discussion}. Finally, we summarise and conclude in Section \ref{sec:conclusion}.

\section{Data}\label{sec:data}

\subsection{FRB Host Galaxy Targets}\label{subsec:frbtargets}

The FRBs selected for this study are taken from published datasets from the CRAFT survey \citep[Commensal Real-time ASKAP Fast Transients;][]{CRAFT,Bannister2019,Shannon2024}, the CHIME/FRB Collaboration \citep{ChimeC1,Amiri2025}, and the MeerTRAP survey \citep[More TRAnsients and Pulsars;][]{MeerTRAP,Rajwade2022,Driessen2024}.  

We place two criteria on targets for inclusion in this study. The first is that the host association is robust, which is evaluated through the PATH framework \citep[Probabilistic Association of Transients to their Hosts;][]{Aggarwal2021}. PATH employs a Bayesian approach to calculate the probability a given transient event is associated with a galaxy, based on the event's localisation and the galaxy's magnitude and size. Following the convention used in \citet{Bhandari2022,Gordon2023}, we consider PATH posterior probabilities P(O|$x$) $\geq$ 0.9 to be robust host associations. 

Our second criterion is that the redshifts of the target host galaxies are $\leq 0.1$. This is due to both the \red{weak intensity} of the \hi\ line, and also the prevalence of contaminating radio frequency interference (RFI) below 1300 MHz ($z\sim$0.092), largely due to the Global Navigation Satellite System (GNSS).

Based on these cuts, our sample consists of nine CHIME FRBs, \red{six} CRAFT FRBs, and two MeerTRAP FRBs. All but one CHIME target (FRB 20200223B) are apparently non-repeating bursts (``non-repeaters" herein), and one of the CRAFT targets (FRB 20211127I) is the FRB previously analysed in \citet{Glowacki_wallaby}. The details of their host galaxies are presented in Table \ref{frbhostoverview}.

\begin{table*}[t]
\begin{threeparttable}

\caption{FRB host galaxies analysed in this study. Repeaters are shown with a dagger.}
\label{frbhostoverview}
\begin{tabular}{llllllll}
\toprule
\headrow FRB Name & Dataset & Host Galaxy & R.A. ($\degree$) & Dec. ($\degree$) & Redshift & Reference \\
\midrule
FRB 20181220A & CHIME & 2MFGC 17440 & 348.6982 & 48.3421 & 0.02746  & \citet{chimefrbs} \\
\midrule
FRB 20181223C & CHIME & SDSS J120340.98+273251.4 & 180.9207 & 27.5476 & 0.03024  & \citet{chimefrbs}\\
\midrule
FRB 20190418A & CHIME & SDSS J042314.96+160425.6 & 65.8123 & 16.0738 & 0.07132  & \citet{chimefrbs}\\
\midrule
FRB 20190425A & CHIME & UGC 10667 & 255.6625 & 21.5767 & 0.03122   & \citet{chimefrbs}\\
\midrule
FRB 20200223B$^\dagger$ & CHIME & SDSS J003304.68+284952.6 & 8.2695 & 28.8313 & 0.06024  & \citet{Ibik2023}\\
\midrule
FRB 20200723B & CHIME & NGC 4602 & 190.1538 & -5.1328 & 0.0085  & \citet{frb20200723b}\\
\midrule
FRB 20201123A & MeerTRAP & J173438.35-504550.4 & 263.6596 & -50.7641 & 0.0507  & \citet{Rajwade2022}\\
\midrule
FRB 20210405I & MeerTRAP & 2MASS J1701249-4932475 & 255.3537 & -49.5465 & 0.0656  &  \citet{frb20210405}\\
\midrule
FRB 20211127I & CRAFT & WALLABY J131913–185018 & 199.8082 & -18.8378 & 0.04695  &  \citet{Glowacki_wallaby}\\
\midrule
FRB 20211212A & CRAFT & SDSS J102924.22+012139.2  & 157.3509 & 1.3608 & 0.0715 & \citet{Shannon2024}\\
\midrule
FRB 20231229A & CHIME & UGC 1234 & 26.4658 & 35.1110 & 0.0190 & \citet{Amiri2025}\\
\midrule
FRB 20231230A & CHIME & J045109.39+022205.02 & 72.7887 & 2.3681 & 0.0298 & \citet{Amiri2025}\\
\midrule
FRB 20240201A & CRAFT & WISEA J095937.44+140519.4 & 149.9060 & 14.0887 & 0.04273 & \citet{Shannon2024}\\
\midrule
\red{FRB 20240210A} & CRAFT & WISEA J003506.47–281619.1 & 8.7768 & -28.2723 & 0.02369 & \citet{Shannon2024} \\
\midrule
\red{FRB 20240312D} & CRAFT & WISEA J032629.50-543557.2 & 51.6227 & -54.5993 & 0.04973 & \citet{Gordon2025} \\
\midrule
\red{FRB 20240615B} & CRAFT & WISEA J021551.26-143729.8 & 33.9633 & -14.6247 & 0.0730 & \citet{Gordon2025} \\
\midrule
FRB 20250316A & CHIME & NGC 4141 & 182.4472 & 58.8492 & 0.00635  & \citet{Amiri2025}\\
\bottomrule

\end{tabular}
\end{threeparttable}
\end{table*}

\subsection{\hi\ Observations \& Processing}\label{subsec:processing}

We first searched for archival \hi\ datasets, and found observations of the hosts of FRB 20231229A and FRB 20250316A \citep{Amiri2025}. Both galaxies have been observed in \hi\ twice; we take the most recent spectra available in each case, corresponding to an ALFALFA \citep[Arecibo Legacy Fast ALFA;][]{Haynes2018} survey observation and an Effelsberg 100-m observation \citep{Haynes1999} respectively. Both of these are at spectral resolutions of 11 km s$^{-1}$, with single-dish beams encompassing several times the angular size of the galaxy. 

A search of the MeerKAT archive hosted by the South African Radio Astronomical Observatory (SARAO) revealed that the coordinates of three targets were contained in archival $L$ band observations. The host of FRB 20200723B \citep{frb20200723b} was the target of a 0.5 hour 4K channelisation (corresponding to $\sim$44 \kms resolution) pointing from "Part 2: 1.28 GHz MeerKAT survey of 12MGS Galaxies in the Southern Hemisphere",  project ID SCI-20220822-LM-01, capture block ID 1671245181. The host of FRB 20201123A \citep{Rajwade2022} was contained in a 1 hour 4K channelisation pointing by the "Legacy Survey of the Galactic Plane", project ID SSV-20180505-FC-01, capture block ID 1610418665, a distance of 0.4$\degree$ from the phase centre. The host of FRB 20231230A \citep{Amiri2025} was contained in a 7.5 hour 4K channelisation pointing from the proposal "Observation of the galaxy cluster A520", project ID SCI-20210212-SG-01, capture block ID 1638727749, a distance of 0.9$\degree$ from the phase centre.

The remaining targets had no other archival data available, and thus we pursued targeted observation with the upgraded Giant Metrewave Radio Telescope \citep[uGMRT;][]{GMRT,uGMRT} and the Five-hundred-meter Aperture Spherical Telescope \citep[FAST;][]{FAST} for northern hemisphere targets, and MeerKAT for southern hemisphere targets. We present summaries of all \hi\ observations in Table \ref{hi_obs}.

\begin{table*}[t]
\begin{threeparttable}

\caption{Summary of FRB host galaxy \hi\ observations.}
\label{hi_obs}
\begin{tabular}{l|cccccc}
\toprule
\headrow FRB & Facility & Observation Date(s) & Channel Width (kHz)\tnote{a} & Time On Source (mins) & Bandpass Calibrator & Gain Calibrator \\
\midrule
 
FRB 20181220A & GMRT & 17/19\textsuperscript{th} August 2024 & 24.414 & 996 & 3C286 & 2202+422\\ 
 & FAST & 30\textsuperscript{th} September 2024 & 7.629 & 10 & - & - \\
\midrule

FRB 20181223C & GMRT & 17/19/23\textsuperscript{rd} August 2024 & 24.414 & 1038 & 3C147 & 1158+248 \\ 
& FAST & 21\textsuperscript{st} Decemeber 2024 & 7.629 & 51 & - & - \\
\midrule

FRB 20190418A & GMRT & 16/17/21\textsuperscript{st} August 2024 & 48.828 & 1056\tnote{b} & 3C48 & 0421+206 \\ 
\midrule 

FRB 20190425A & GMRT & 22/26/27\textsuperscript{th} August 2024 & 24.414 & 990 & 3C286 & 1640+123 \\ 
& FAST & 8$^\text{th}$ February 2025 & 7.629 & 3 & - & - \\
\midrule

FRB 20200223B & GMRT & 10/18/23\textsuperscript{rd} November 2023 & 48.828 & 1020 & 3C48 & 0029+349 \\ 
& FAST & 1\textsuperscript{st} October 2024 & 7.629 & 1 & - & -\\
\midrule

FRB 20200723B & MeerKAT & 17\textsuperscript{th} December 2022 & 208.984 & 30 & J1939-6342 & J1215-1731\\ 
\midrule

FRB 20201123A & MeerKAT & 12\textsuperscript{th} January 2021 & 208.984 & 60 & J1939-6342 & J1744-5144\\ 
\midrule

FRB 20210405I & MeerKAT & 21\textsuperscript{st} June 2024 & 13.062 & 240 & J1939-6342 & J1744-5144\\ 
\midrule

FRB 20211127I & MeerKAT & 18\textsuperscript{th} March 2023 & 3.265 & 180 & J1939-6342 & J1311-2216 \\  
\midrule

FRB 20211212A & MeerKAT & 13\textsuperscript{th} December 2022 & 3.265 & 180 & J0408-6545 & J1008+0730 \\ 
\midrule

FRB 20231229A & Arecibo\tnote{c} & - & 24.390 & {$\sim$1} & - & -\\ 
\midrule

FRB 20231230A & MeerKAT & 5\textsuperscript{th} December 2021 & 208.984 & 450 & J0408-6545 & J0503+0203 \\ 
\midrule

FRB 20240201A & MeerKAT & 21\textsuperscript{st} February 2025 & 26.123 & 240 & J0408-6545 & J1008+0730 \\ 
& FAST & 13\textsuperscript{th} November 2024 & 7.629 & 23 & - & -\\
\midrule

{FRB 20240210A} & MeerKAT & 28\textsuperscript{th} July 2025 & 26.123 & 240 & J0408-6545 & J0025-2602 \\ 
\midrule

{FRB 20240312D} & MeerKAT & 29\textsuperscript{th} June 2025 & 26.123 & 240 & J0408-6545 & J0303-6211 \\ 
\midrule

{FRB 20240615B} & MeerKAT & 3\textsuperscript{rd} August 2025 & 26.123 & 240 & J0408-6545 & J0240-2309 \\ 
\midrule

FRB 20250316A & Effelsberg\tnote{d} & {Oct/Nov 1989} & 11.965 & - & - & - \\ 
\bottomrule
\end{tabular}

\begin{tablenotes}[hang]
\item[a]Raw observation channel widths; we rebin to lower resolution in most cases.
\item[b]16$^\text{th}$ August observation is corrupted, true on source time is $\sim$66\% of this.
\item[c]\citet{Haynes2018} - \red{ALFALFA survey was conducted over a 7 year time span with an average integration time of 48 seconds.} 
\item[d]\citet{Haynes1999} - \red{precise on source time unknown.}
\end{tablenotes}

\end{threeparttable}
\end{table*}

\subsubsection{GMRT Observations and Data Reduction}\label{subsubsec:gmrt}

Through proposals 45$\_$059 and 46$\_$112, we obtained a total of 114 hours of $L$ band observation ($\sim$85 hours on-source) centred on the hosts of FRB 20181220A, FRB 20181223C, FRB 20190418A, FRB 20190425A \citep{chimefrbs} and FRB 20200223B \citep{Ibik2023}. These observations were taken in November 2023 and August 2024, using the GMRT Wideband Backend as the correlator. Depending on the location of the \hi\ line in the band, we used 100 MHz and 200 MHz bandwidths between 1260 - 1460 MHz, covering the entire $z=0 -0.1$ range. These were subdivided into 4096 channels, resulting in velocity resolutions of $\sim6$ km s$^{-1}$ and $\sim$12 km s$^{-1}$ respectively. 

The raw data were transferred and processed using CASA \citep{CASA} on the \texttt{ilifu} supercomputing cloud system. Initial flagging of corrupted antennas and scans was completed manually, and gain, bandpass, and phase calibrations were conducted following standard procedures. The data were rebinned to 1 MHz for self-calibration, and the calibration solutions were applied to the visibilities around the \hi\ line region. Final deep continuum images were made out to the first null of the uGMRT primary beam, whose resulting  model continuum visibilities were subtracted from the corrected visibilities using the CASA task \textsc{uvsub}. The residual data were then run through a final round of automatic RFI removal. 

A uv-plane continuum subtraction was completed with the CASA task \textsc{uvcontsub}, excluding 500 \kms\ around the region of the \hi\ line with a first order polynomial fit. We then made data cubes surrounding the host galaxy coordinates using the CASA task \textsc{tclean} with Briggs weighting. Various robust levels and resolutions were employed to test different emphases on sensitivity and structure, ensuring that around 5 pixels covered the beam. Due to some residual variations in the baseline, in a few cases we ran a final image-plane based continuum subtraction using the CASA task \textsc{imcontsub}, excluding the channels visibly containing \hi\ emission with a second order polynomial. Finally, each spectral-line datacube was \red{corrected for its frequency-dependent primary beam shape and} convolved to its largest synthesised common beam.  

\subsubsection{MeerKAT Observations and Data Reduction}\label{subsubsec:meerkat}

Through three MeerKAT proposals (project IDs SCI-20220822-MG-01, SCI-20230907-MG-02 and SCI-20241103-MG-01), we obtained a total of 14 on-source hours, centred on the hosts of \red{seven FRBs}. The observations were conducted from \red{December 2022, through to August 2025 (capture block IDs 1670970986, 1679181682, 1718998942, 1739386395, 1740164485, 1751156931, 1753661382, and 1754260765)}. Each observation was taken in the $L$ band (107 MHz centred at 1358.5 MHz) at 32K and 32K-zoom spectral resolutions, which corresponds to velocity resolutions of 5.5 km s$^{-1}$ and 0.7 km s$^{-1}$ respectively.

The data were stored on \texttt{ilifu} and processed using \texttt{PROCE\-SSMEERKAT}\footnote{https://idia-pipelines.github.io/docs/processMeerKAT}, a dedicated MeerKAT reduction pipeline which also utilises CASA. We rebinned the 32K-zoom data to a maximum of 64 K mode (i.e. 13.062 kHz wide channels) for this step. 

Model continuum visibilities were subtracted from the corrected data using \textsc{uvsub}, and then a further \textsc{uvcontsub} step was run. Spectral-line cubes were made using \textsc{tclean} with Briggs weighting, again with various robust levels, resolutions, and cleaning depths. Each cube was primary beam corrected using  \texttt{katbeam}\footnote{https://github.com/ska-sa/katbeam} and then was convolved to a synthesised common beam.

\subsubsection{FAST Observations and Data Reduction}\label{subsubsec:fast}

To supplement a few northern hemisphere observations, and in reaction to more recent FRB localisations, we proposed for FAST time through proposal SQB-2024-0174 and were awarded four hours of observation. The targets were observed in On-Off mode with the spec(W+N) backend; this spectral backend records the full 400 MHz (1.05-1.45 GHz) bandwidth with a resolution of 7.629 kHz, and a narrower 31.25 MHz region with a resolution of 476.83 Hz. The On-Off mode utilises the periodic injection of a standardised noise diode for flux calibration. 

The data are stored in FITS format as a dual polarisation time series of bandwidth data. We separated the on and off source data and performed a temperature calibration using the high power noise diode \citep[e.g. Section 3.1 of][]{Jing2023} to convert from raw power to brightness temperature. Due to the electromagnetic wave reflection that is common in single dish instruments, the FAST data contains periodic fluctuations in the bandpass which manifest as standing waves. We removed this signal using a fast Fourier transform (FFT) based approach, modelled off the methods of \citet{Xu2024}. Using the spectroscopic redshifts of each target, we extracted 2000 km s$^{-1}$ around the \hi\ line and stacked each observation weighted by its RMS noise. 

To subtract the baseline, we performed a polynomial fit with 250 km s$^{-1}$ excluded around the \hi\ frequency. The polynomial was chosen as either 3\textsuperscript{rd} or 4\textsuperscript{th} degree such that the resulting baseline was flat around the line. In a few datasets, the baseline was too erratic for a clean subtraction; for these cases we first subtracted the off source summed spectra from the on source summed spectra, and then performed the baseline polynomial fitting. This was completed separately for each polarisation, and the final spectra was obtained by stacking them together.

\section{\hi\ Results}\label{sec:results}

\begin{figure*}[!t]
    \centering
    \includegraphics[width=\linewidth]{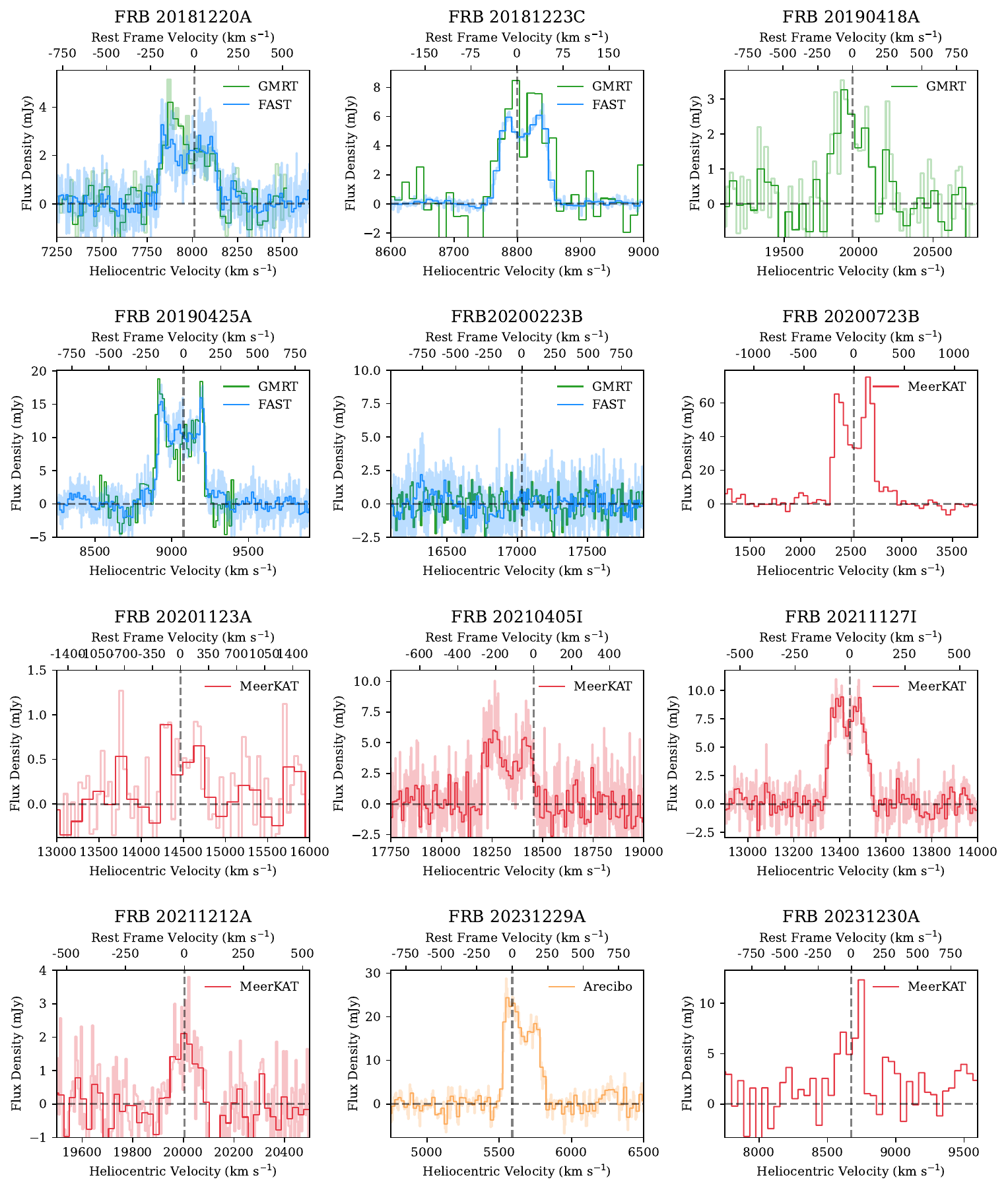}
    \caption{\hi\ spectra of the FRB host galaxies. The red, green, and blue colours represent dedicated observations from FAST, GMRT, and MeerKAT, with the two orange spectra taken from archival data. When necessary, the data are binned by various channel widths; fainter lines in the plots represent the raw data, with bolder lines representing the binned data. The vertical dashed line represents the rest frame of the hosts', determined by their optical redshifts.}
    \label{fig:spectra}
\end{figure*}

\begin{figure*}[!t]
\ContinuedFloat
    \centering
    \includegraphics[width=\linewidth]{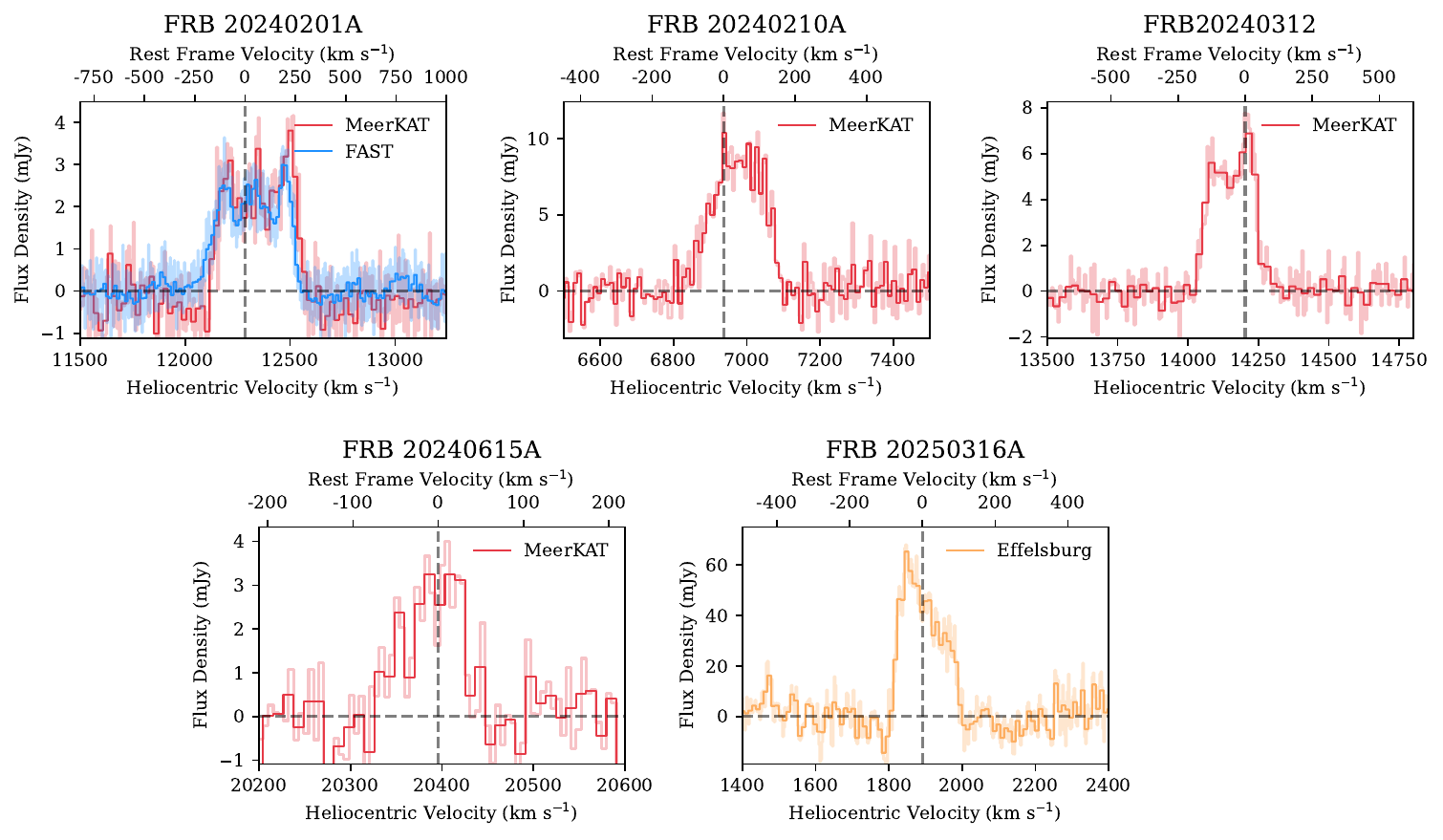}
    \caption{FRB host galaxy \hi\ spectra (continued)}.
    \label{fig:spectra}
\end{figure*}

\begin{figure*}[!t]
    \centering
    \includegraphics[width=0.93\linewidth]{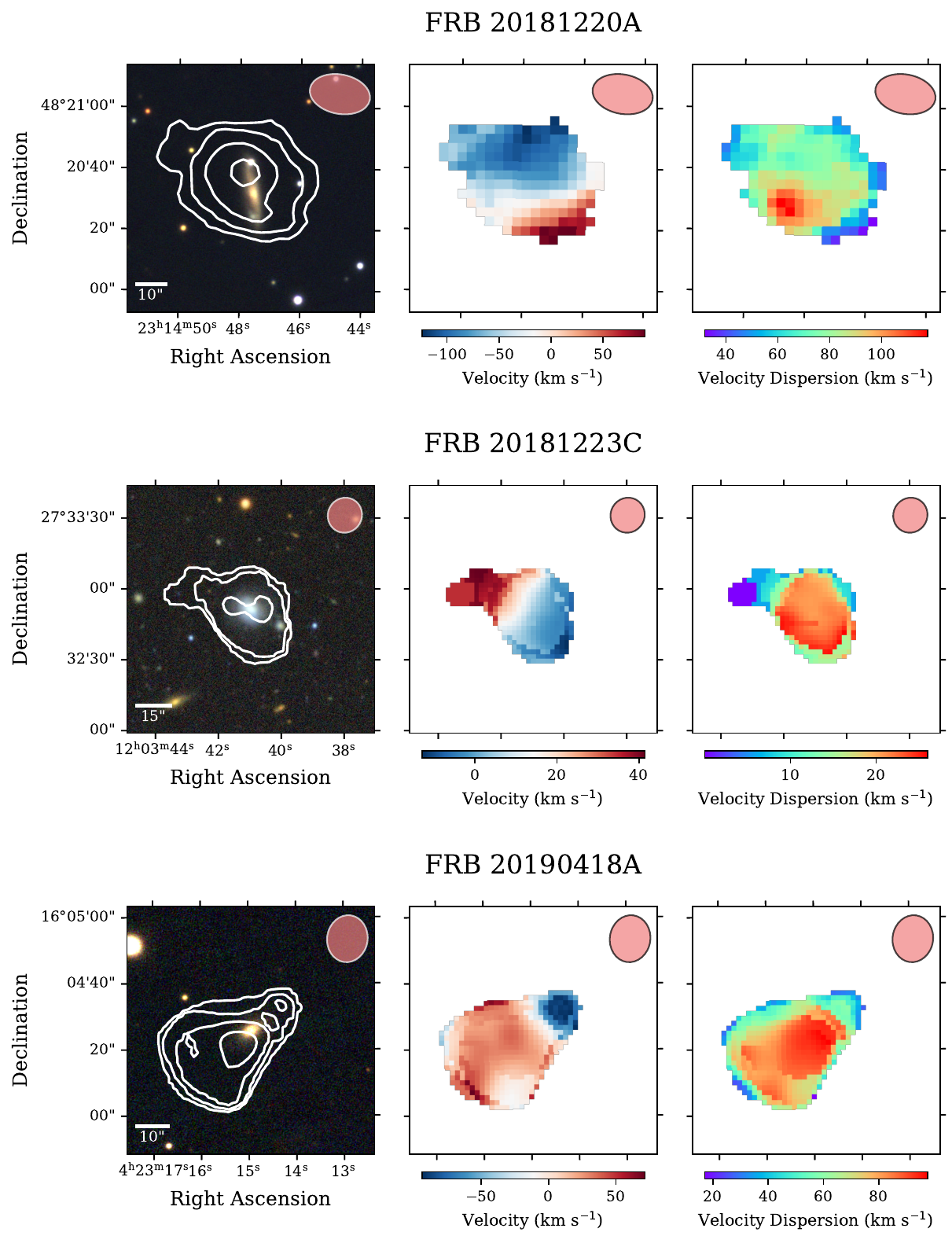}
    \caption{FRB host galaxy moment maps, displaying total intensity (left), velocity (centre) and velocity dispersion (right). The lowest contours are at the 3$\sigma$ level, with the higher contours set at varying multiples of that level. When the FRB's localisation region is significantly smaller than the size of its host, we include its position shown in magenta in the intensity maps and black in the others; a star represents a localisation region too small to be shown, whereas a cross and dashed ellipse shows the estimated position and 1$\sigma$ uncertainty region. The beam sizes are shown in the upper right hand corner. Velocities are displayed in the rest frame, defined with respect to the optical redshift.}
    \label{fig:momentmaps}
\end{figure*}

\begin{figure*}[!t]
\ContinuedFloat
    \centering
    \includegraphics[width=0.97\linewidth]{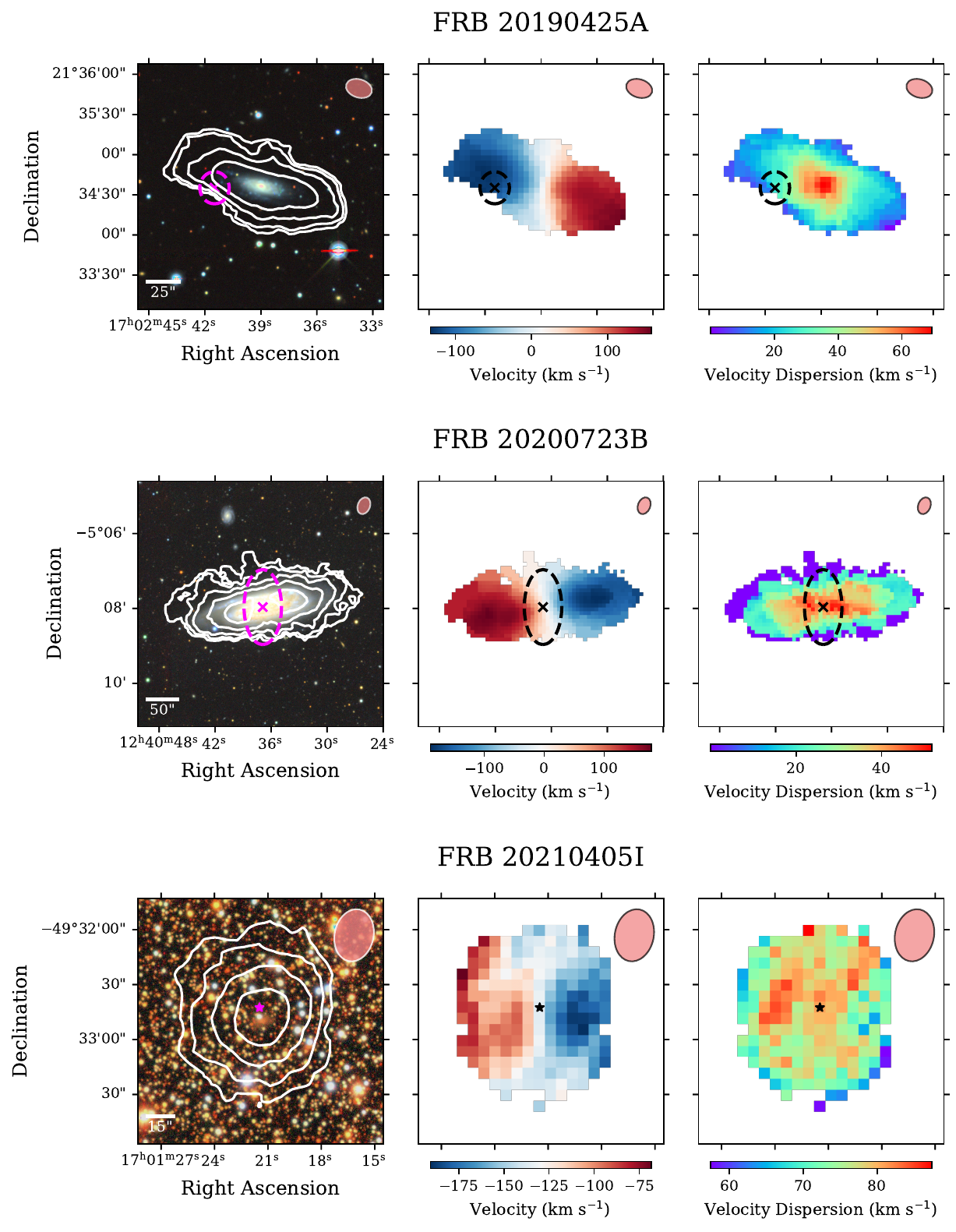}
    \caption{FRB host galaxy moment maps (continued).}
\end{figure*}

\begin{figure*}[!t]
\ContinuedFloat
    \centering
    \includegraphics[width=0.97\linewidth]{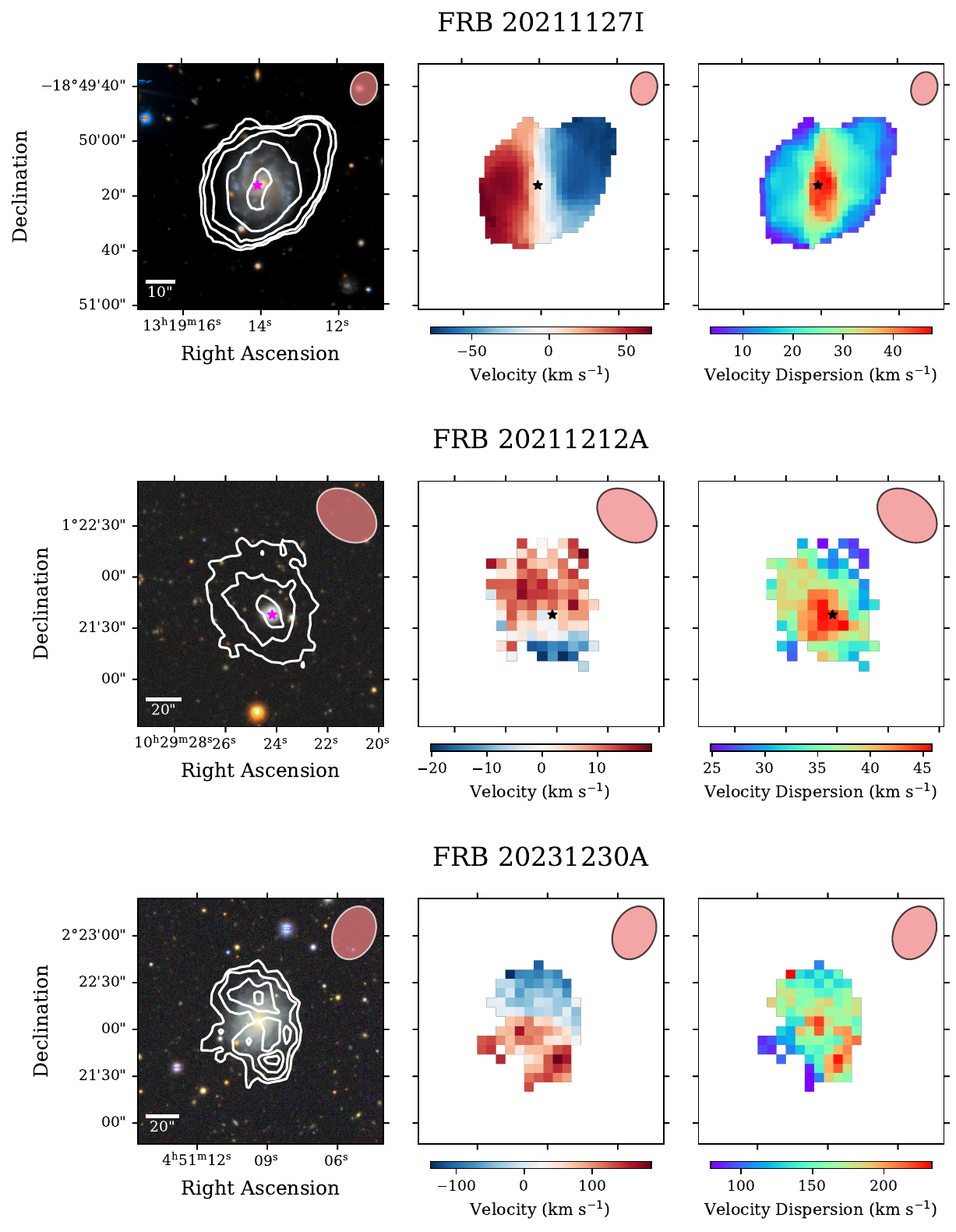}
    \caption{FRB host galaxy moment maps (continued).}
\end{figure*}

\begin{figure*}[t!]
\ContinuedFloat
    \centering
    \includegraphics[width=0.97\linewidth]{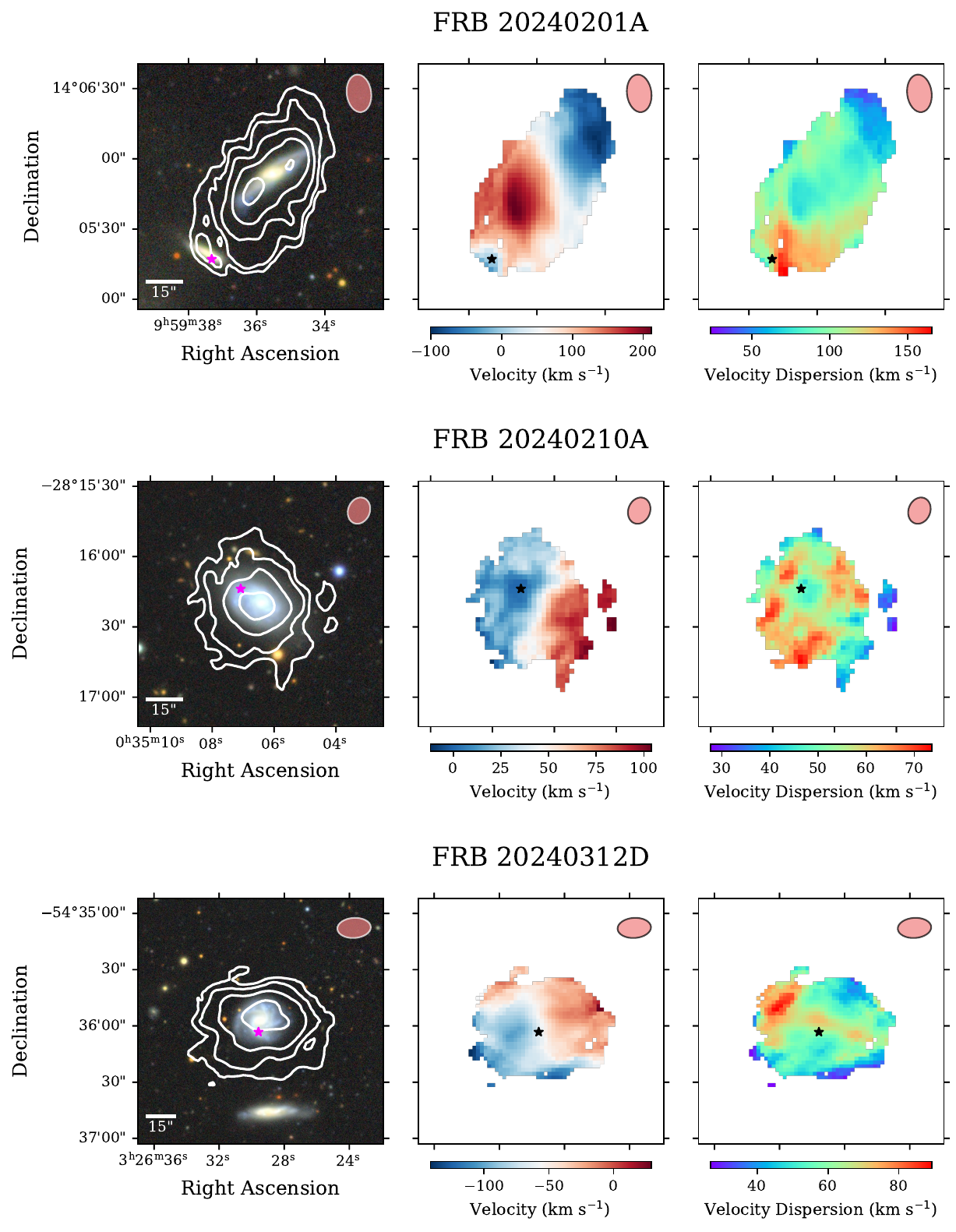}
    \caption{FRB host galaxy moment maps (continued).}
\end{figure*}

\begin{figure*}[t!]
\ContinuedFloat
    \centering
    \includegraphics[width=0.97\linewidth]{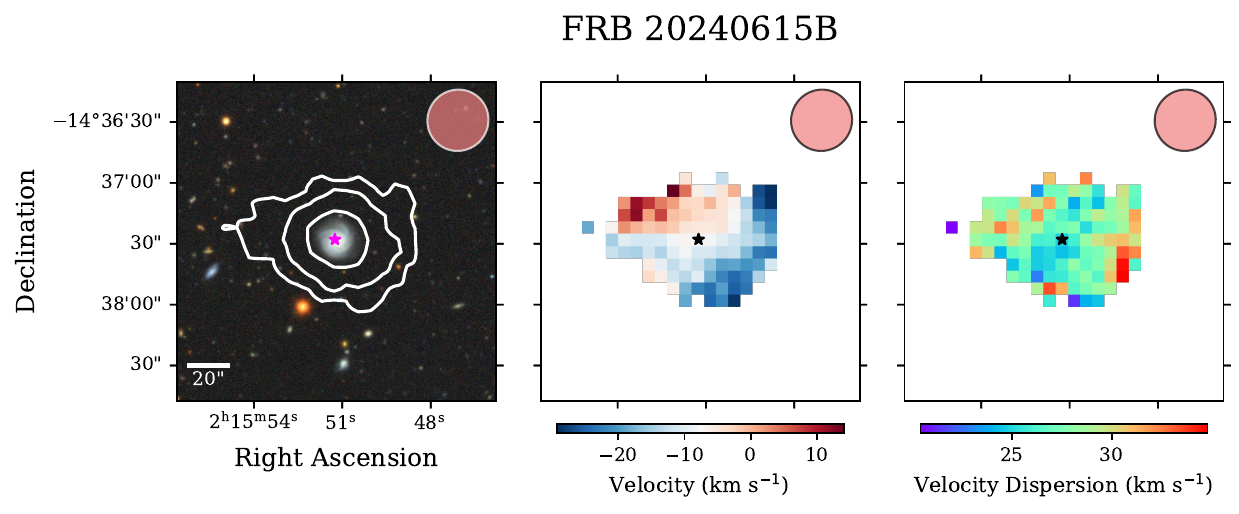}
    \caption{FRB host galaxy moment maps (continued).}
\end{figure*}

\begin{table*}[th!]
\caption{Host galaxies properties derived from radio observations: \hi\ flux, \hi\ mass, line width (defined by the width between the 50\% peak flux levels), radio continuum flux density, radio-derived star formation rate, \hi\ disk radius, and virial mass.}
\label{tab:radioproperties}
\begin{threeparttable}
\begin{tabular}{lccccccccc}
\toprule
\headrow FRB Name & Observation & \shi & $\log$(\mhi)& $W50$  & $S$\textsubscript{cont}\tnote{a} & SFR\textsubscript{radio} & \rhi & $\log$($M_{200}$/\msun)\tnote{b} \\
\headrow & & (Jy \kms) & (\msun) &  (\kms) & (mJy) & (\msun \yr) & (kpc) & \\
\midrule

FRB 20181220A & GMRT & 0.76 \PM\ 0.12 & 9.44 \PM\ 0.08 & 126 \PM\ 14 & 2.72 & 2.29$^{+1.30}_{-0.83}$  & 20.5 \PM\ 10.0 & 11.0 \\
& FAST & 0.72 \PM\ 0.08 & 9.42 \PM\ 0.05 & 315 \PM\ 2 & N/A & N/A & N/A & N/A \\

\midrule

FRB 20181223C & GMRT & 0.60 \PM\ 0.10 & 9.42 \PM\ 0.08 & 70 \PM\ 5 & $\leq$0.08 & $\leq$0.15  & 11.1 \PM\ 2.4 & 10.6 \\
& FAST & 0.45 \PM\ 0.04 & 9.30 \PM\ 0.05 & 79 \PM\ 1 & N/A  & N/A & N/A & N/A \\

\midrule

FRB 20190418A & GMRT & 0.82 \PM\ 0.15 & 10.33 \PM\ 0.09 & 217 \PM\ 19 & $\leq$0.06  & $\leq$ 0.51  & 36.2 \PM\ 8.2 & - \\
\midrule

FRB 20190425A & GMRT & 3.55 \PM\ 0.39 & 10.22 \PM\ 0.05 & 303 \PM\ 3 & 1.11 & 1.35$^{+0.75}_{-0.48}$  & 35.0 \PM\ 5.0 & 12.0 \\
& FAST & 3.61 \PM\ 0.36 & 10.23 \PM\ 0.05 & 307 \PM\ 1 & N/A & N/A & N/A & N/A \\

\midrule

FRB 20200223B & GMRT & $\leq$0.13  & $\leq$9.377 & N/A & 0.10  & 0.59$^{+0.32}_{-0.21}$ & N/A & N/A\\
& FAST & $\leq$ 0.14 & $\leq$ 9.394 & N/A & N/A & N/A & N/A & N/A\\
\midrule

FRB 20200723B & MeerKAT & 20.47 \PM\ 2.11 & 9.84 \PM\ 0.05 & 401 \PM\ 4  & 47.7 & 3.42$^{+1.97}_{-1.25}$  & 23.0 \PM\ 2.9 & 12.0 \\
\midrule

FRB 20201123A & MeerKAT & 0.41 \PM\ 0.18 & 9.72 \PM\ 0.16 & 415 \PM\ 36 & 5.15 & 10.9$^{6.50}_{-4.08}$ & N/A &  N/A \\
\midrule 

FRB 20210405I & MeerKAT & 1.00 \PM\ 0.12 & 10.34 \PM\ 0.06 & 245 \PM\ 5 & 0.59  & 2.85$^{+1.63}_{-1.04}$  & 29.3 \PM\ 3.3 & 12.7 \\
\midrule

FRB 20211127I & MeerKAT & 0.90 \PM\ 0.10 & 9.99 \PM\ 0.05 & 158 \PM\ 3  & 1.16  & 2.82$^{+1.61}_{-1.02}$  & 14.5 \PM\ 2.5 & 11.7 \\
\midrule

FRB 20211212A & MeerKAT & 0.17 \PM\ 0.04 & 9.65 \PM\ 0.11 & 87 \PM\ 8 & 0.31 & 1.92$^{+1.09}_{-0.69}$ & 25.9 \PM 9.0 & - \\
\midrule 

FRB 20231229A & Arecibo & 5.29 \PM\ 0.54 & 9.96 \PM\ 0.05 & 256 \PM\ 3 & N/A & N/A & N/A & N/A\\
\midrule 

FRB 20231230A & MeerKAT & 0.18 \PM\ 0.04 & 8.89 \PM\ 0.09 & 206 \PM\ 21 & $\leq$5.19  & $\leq$4.47  & 12.3 \PM\ 3.4 & - \\
\midrule 

FRB 20240201A\tnote{$\dagger$} & MeerKAT & 1.01 \PM\ 0.11 & 9.96 \PM\ 0.05 & 373 \PM\ 5 & 0.42  & 1.04$^{+0.57}_{-0.37}$  & - & - \\
& FAST & 0.89 \PM\ 0.09 & 9.90 \PM\ 0.05 & 384 \PM\ 1 & N/A & N/A & N/A & N/A \\
\midrule 

\red{FRB 20240210A} & MeerKAT & 1.51 \PM\ 0.16 & 9.61 \PM\ 0.05 & 166 \PM\ 1 & 1.67  & 1.19$^{+0.66}_{-0.43}$ & 16.8 \PM\ 4.2 & 11.38 \\
\midrule 

\red{FRB 20240312D} & MeerKAT & 1.25 \PM\ 0.13 & 10.19 \PM\ 0.05 & 182 \PM\ 1 & 1.26 & 3.31$^{1.90}_{1.21}$ & 27.2 \PM\ 4.3 & - \\
\midrule 

\red{FRB 20240615B} & MeerKAT & 0.25 \PM\ 0.03  & 9.84 \PM\ 0.06 & 76 \PM\ 1 & 0.30  & 1.99$^{+1.12}_{-0.72}$ & 30.3 \PM\ 3.0 & 11.39 \\
\midrule 

FRB 20250316A & Effelsberg & 7.13 \PM\ 0.79 & 9.13 \PM\ 0.05 & 132 \PM\ 2 & N/A & N/A & N/A & N/A\\
\bottomrule
\end{tabular}
\begin{tablenotes}[hang]
\item[a]Estimates are accompanied by 10\%  uncertainties due to absolute flux scaling errors.
\item[b]Dashes indicate hosts with inclinations too near 0 for proper $V_\text{rot}$ correction.
\item[$\dagger$]\hi\ estimates include entangled neighbour emission.
\end{tablenotes}
\end{threeparttable}
\end{table*}

Here we highlight the global \hi\ properties of the sample. For each spatially resolved observation (i.e., the GMRT and MeerKAT datacubes), we use the SoFiA-2 \citep{Westmeier2021} source finding pipeline to extract \hi\ sources and generate moment maps. We use the pipeline's default smooth-and-clip (S+C) finder with parameters following those used in \citet{Westmeier2022} for the pilot data release of the WALLABY survey (which SoFiA-2 was designed for). This includes a  \texttt{scfind.th\-reshold} value of 3.5 and a \texttt{reliability.minSNR} value of 3, to ensure only robust signals are picked up. For detections with particularly narrow line widths, we rerun SoFiA-2 with a maximum spectral axis smoothing kernel equal to this apparent width.

Figure~\ref{fig:spectra} presents spectra for each target, binned to lower spectral resolution where sensible\footnote{See Table \ref{tab:columndensity} in Appendix for more information.\label{sharedfn}}. The GMRT and MeerKAT spectra are generated by summing the flux from pixels included in a 2D mask, which itself is a flattened projection of the 3D mask generated by SoFiA-2. We also present moment 0 (intensity), moment 1 (velocity), and moment 2 (velocity dispersion) maps for these targets in Figure~\ref{fig:momentmaps}, overlaid on imaging from the DESI Legacy Imaging Survey \citep{DESI}, the Dark Energy Camera Plane Survey \citep[DECaPS]{DECaPS}, and the Pan-STARRs1 survey \citep{PanSTARRS}. The lowest contours are defined as the 3$\sigma$ level measured by SoFiA-2\footnotemark[\value{footnote}].

Table~\ref{tab:radioproperties} presents a summary of the global \hi\ properties of the sample. These properties are measured from the masked spectra; i.e. only voxels in the 3D SoFiA-2 mask are included, and as a result they are slightly different to those displayed in Fig.~\ref{fig:spectra}. We calculate \hi\ masses following \citet[equation 48]{Meyer2017}, assuming a 10\%\ absolute flux scale uncertainty on all measurements due to calibration \red{alongside the noise contribution}. As radio continuum (RC) emission is considered a robust tracer of star formation due to its association with young, massive stars, when RC sources are present, we estimate an SFR following the methods of \citet[equation 22]{Molnar2021}. 

We estimate the radius of the \hi\ emission, \rhi, out to the 3$\sigma$ contour level by measuring the average surface density\footnotemark[3] of the intensity maps as a function of elliptical radius from a chosen centre, which we choose to be the optical centre of brightness in the survey imaging. This \rhi\ estimate is \red{technically a lower limit; as many of our maps do not reach the nominal 1 \msun pc$^{-2}$ threshold used in the literature, they are not} necessarily half the disk size $D_\text{HI}$ commonly used to describe HI disks e.g. for the mass-size relation \citep{Wang2016}.  We estimate the ellipticity and position angles of the galaxies using 2MASS \citep{2MASS} K-band data where possible; otherwise we fit elliptical isophotes to the \red{legacy photometry presented in Fig.~\ref{fig:momentmaps}}. These values are then combined with estimates of the galaxies' rotation velocities $V_\text{rot}$ to make broad approximations of the dynamical mass $M_\text{dyn}$ and hence the virial mass $M_{200}$ for these galaxies. We estimate $V_\text{rot}$ using 

\[ V_\text{rot} = \frac{W_{50}}{2\sin i} \]

\noindent where $W_{50}$ is the line width at 50\%\ of the peak flux and $i$ is the inclination of the galaxy; we discard four galaxies with inclinations lower than 10$\degree$. To ensure these rotational velocities are in the flat part of each galaxies' rotation curve, we examine position-velocity diagrams across the major axes and find that all targets extend beyond the break. We do not apply systematic corrections for instrumental broadening and turbulence as we expect such corrections to be small for \red{such} high resolution data, and we are not attempting a fully robust measurement. We calculate $M_\text{dyn}$ and $M_{200}$ following \citet[equations 10 \& 13]{Yu2020}. 

\subsection{Marginal / Non-Detections}\label{subsec:nondet}

\begin{figure}[t!]
    \centering
    \hspace*{-1cm}
    \includegraphics[width=0.85\linewidth]{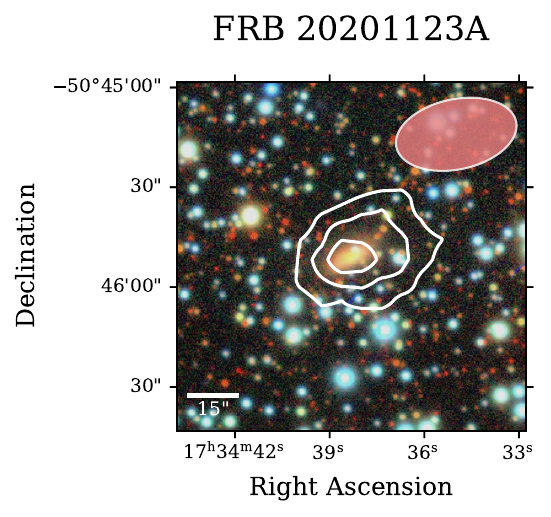}
    \caption{Intensity map of the unresolved detection of the host of FRB 20201123A.}
    \label{fig:solomap}
\end{figure}

As seen in Fig.~\ref{fig:spectra}, the host of FRB 20201123A peaks very marginally above the noise. This host is at a relatively high redshift of 0.0507, and the MeerKAT observation was 0.4$\degree$ from the phase centre. Due to this, and the beam size, we presume the picked-up signal is simply the unresolved core emission; as such, we do not place much weight on its derived \hi\ properties, considering its \hi\ mass to be a lower limit. Only its intensity map is displayed \red{in Fig.~\ref{fig:solomap}}, and we do not attempt to measure its asymmetry.   

% It is represented by the fainter upward facing red arrow in Fig.~\ref{fig:hi_properties}. 

Fig.~\ref{fig:spectra} also shows the non-detection of the host of FRB 20200223B. 
We place a 5$\sigma$ upper limit on its \hi\ mass using both the GMRT and FAST data, assuming a linewidth of 200 \kms\ and a disk size of 30 kpc. As outlined in \citet{Ibik2023}, this host is in a transitional regime between star-formation and quiescence, which reflects its lower gas fraction. This is uncommon for FRB hosts; \citet{Gordon2023,Sharma2024} clearly find a preference towards star-forming galaxies, a conclusion well supported by our findings given that \red{16 of our 17} targets are emitting in \hi. 

% This leaves us with 12 galaxies with which to evaluate the trend of asymmetry in FRB hosts.

% This non-detection is designated by the fainter downward facing red arrow in Fig.~\ref{fig:hi_properties}, which shows that such an upper limit does not necessarily place it in a particularly unusual position with respect to the background population, but does show it is slightly on the fringe of the FRB host population. This would be further highlighted if we were to use a 3$\sigma$ limit. 

\section{Asymmetry Calculations}\label{sec:asym}

To fully investigate the degree of disturbance in the \hi\ distributions of FRB hosts, we employ close by-eye examination of the spectral profiles, the moment maps, and individual channel intensity maps. However, we also wish to quantify any disturbance to allow for direct comparison with background galaxies. In this section, we cover our methods for quantifying and calculating the asymmetry of the \hi\ detections. 

\subsection{Asymmetry Metrics}\label{subsec:asym_metrics}

There are a number of common metrics used in the literature to assess asymmetry \citep[see the introduction of][for an overview]{Deg2023}; in this study we utilise five: \Aflux, \Aspec, \AoD, $A_\text{2D}$, and \AtD. The first is a common historical measurement used to consider the lopsidedness of a spectrum, and is defined by

\begin{equation}
A_\text{flux} = \left|\ \text{max}(F_\text{l}/F_\text{u}\ ,\ F_\text{u} /F_\text{l})\ \right|
\end{equation}

\noindent where 

\begin{equation}
F_\text{l} = \int_{v_1}^{v_\text{sys}} F_{v}\ dv
\end{equation}

\noindent is the integrated flux of the lower half of the spectrum up to some central velocity $v_\text{sys}$, and  

\begin{equation}
F_\text{u} = \int_{v_\text{sys}}^{v_2} F_v\ dv
\end{equation}

\noindent is the integrated flux of the upper half. This central velocity is generally defined as the systemic velocity, given by the midpoint of the spectrum at the 20\% flux level.

The second metric, \Aspec, was introduced by \citet{Reynolds2020LVHIS} to emphasise more local disturbances through investigation of channel-by-channel variation, defined by 

\begin{equation}
A_\text{spec} = \frac{\sum_{i=1}|\ S(i)-S(i_{\text{flip}})\ |}{\sum_{i=1}|\ S(i)\ |}
\end{equation}

\noindent where $S(i)$ and $S(i_\text{flip})$ are the fluxes of the $i^\text{th}$ channel and its mirrored channel pair about a central channel. This centre is generally different than $v_\text{sys}$, and is instead chosen as the flux-weighted mean velocity to highlight the differences in profile far away from the centre of mass, e.g. through an extended tail on one side of the spectrum. These metrics were used in \citet{Michalowski21,Glowacki_wallaby} to compare FRB hosts with wider LVHIS and HALOGAS \hi\ survey populations \citep{Reynolds2020LVHIS}, and thus we include calculations here. 

Recently, \citet{Deg2023} introduced new metrics to properly account for the bias contributed by noise to the asymmetry of a profile, and to take advantage of the full information that three-dimensional datacubes can offer. As next generation surveys, such as WALLABY and LADUMA, will resolve many \hi\ galaxies, these higher dimensional metrics are likely to be the most robust way to quantify a galaxy's disturbance, especially in low SNR regimes. Thus, we also calculate values for \AoD, $A_\text{2D}$, and \AtD, based on equation 20 in \citet{Deg2023}:

\begin{equation}
A_\text{1D/2D/3D} = \left(\frac{P_\text{sq,m}-B_\text{sq,m}}{Q_\text{sq,m}-B_\text{sq,m}} \right)^{1/2}
\end{equation}

\noindent where 

\begin{equation}
P_\text{sq,m} = \sum_i^N (\ f_i-f_{-i}\ )^2 \end{equation}

\noindent is the sum of the squared difference between each flux channel/pixel/voxel $f_i$ in an \hi\ profile and its mirrored counterpart $f_{-i}$ about a 1D/2D/3D rotation point, 

\begin{equation}
Q_\text{sq,m} = \sum_i^N (\ f_i+f_{-i}\ )^2, \end{equation}

\noindent and $B_\text{sq,m}$ is the contribution of the noise to $P_\text{sq,m}$, which for uncorrelated Gaussian noise $\sigma$ can be approximated as $2N\sigma^2$. These metrics are calculated by the \texttt{3DACs} code\footnote{https://github.com/NateDeg/3DACS}, which simply requires a datacube, a source mask, and a 3D rotation voxel as inputs. 

It should be noted that in some cases, $B_\text{sq,m}$ may be larger than $Q_\text{sq,m}$, which causes the result to be undefined. This indicates that the noise is dominating the measured asymmetry $P_\text{sq,m}$. \texttt{3DACs} returns a value of -1 in these cases, which should be treated essentially equal to a value of zero. 

The choice of rotation voxel is critical to the usefulness and applicability of these metrics. We want to define it such that tidal features cannot introduce bias in either the spatial or spectral dimensions. As such, we choose to first find the optical centre of brightness in the background survey image, and take the flux weighted centre of the spectrum at the closest pixel to this 2D point. This choice is thus sensitive to any gas volume that is away from the centre of the galaxy's core. 

% Unfortunately, this method may be unfeasible for large volume blind surveys such as WALLABY, which may choose to instead take the flux weighted centre of the \hi\ intensity map, or use detailed kinematic modelling to find the dynamic centre. While these choices are certainly valid for measuring the asymmetry in the \hi\ distribution, we suspect that connecting back to the optical disk is a more robust way to describe the asymmetry of a galaxy as a whole. 

We note that as we cannot run \texttt{3DACs} on the single dish spectra, for consistency we calculate all \AoD\ values externally (following the same methodology as above) using the flux weighted centre as the pivot. As \texttt{3DACs} estimates the RMS noise of the 1D spectrum based on an expected scaling of the 3D RMS noise when summing over spatial pixels, rather than measuring the noise itself, our values differ slightly, though never significantly. 

When extracting final values for each of the 1D metrics, we consider the optimal spectral resolution. Each can become temperamental as the SNR of the \hi\ line decreases and as the linewidth (in channels) decreases, and thus we take measurements 
at many different rebinning ratios \red{(i.e. until the spectral line is unresolved)} to estimate the most stable resolution. This estimation differs between target and is done purely by eye; we inspect the variation of each metric to find the resolution which appears coincident with the most overall stable point in the three metrics, aiming to use the highest resolution possible. We highlight an example case in Fig.~\ref{fig:asymdiagnostic} in the Appendix. 

\subsection{Asymmetry Thresholds}\label{subsec:asym_thresh}

We wish to identify the thresholds above which a profile is considered asymmetric for each of our metrics. \citet{Espada2011} modelled the \Aflux\ distribution of AMIGA isolated galaxies, finding a 2$\sigma$ threshold of 1.26.
\citet{Reynolds2020LVHIS} did the same with the low density fields of LVHIS and HALOGAS, and the higher density field of VIVA, finding 2$\sigma$ thresholds of 1.25, 1.17, and 1.53 respectively for galaxies with stellar masses $\geq10^9$ \msun. \citet{Watts2020} focused on the more representative xGASS sample, and chose to define asymmetry by generating noiseless mock spectra with inherent \Aflux\ equal to 1.1, perturbing them by various noise levels, and finding the threshold (as a function of SNR) within which 80\%\ of the spectra are measured. Considering these results and the fact that a number of our targets are low SNR ($<$10), we choose a flat threshold of 1.3; this is similar to the 3$\sigma$ limits of \citet{Espada2011,Reynolds2020LVHIS}, and corresponds to the low SNR threshold of \citet{Watts2020}.

\citet{Reynolds2020LVHIS} also measured \Aspec\ for the LVHIS, HALOGAS, and VIVA samples, finding 2$\sigma$ thresholds of 0.428, 0.263, and 2.031 respectively. As this metric is the most affected by noise, we choose a slightly arbitrary conservative threshold of 0.5; this also reflects the fact that FRB hosts are not \red{exclusively} found in low-density environments.

As the \AoD, $A_\text{2D}$, and \AtD\ measurements are even newer, there are not many comparisons to be made with wider populations from which to infer at what point a profile is considered asymmetric. \citet{Perron-Cormier2025} recently applied these metrics to data from the WALLABY pilot survey \citep{Westmeier2022} and found that an \AtD\ threshold of 0.5 corresponds to highly disturbed galaxies, though this value likely represents the extreme end of the disturbance spectrum. Fig. 8 of \citet{Deg2023} presents a trial analysis of 500 galaxies from a simulated WALLABY-like survey, showing a median visual disturbance of approximately 0.2, with most values falling within 0.1 of this median. Thus, an \AtD\ value of 0.3 can serve as a rough threshold for identifying galaxies with above-average disturbance. In Section ~\ref{sec:individuals}, we also present asymmetry measurements of the disturbed FRB host from \citet{Lee-waddell23} to use as a reference.  

\subsection{Uncertainties}\label{subsec:asym_unc}

The uncertainties in the 1D metrics come from a number of sources; we first consider statistical contributions arising from the inherent noise component of each channel. We estimate this statistical uncertainty following \citet{Michalowski21, Glowacki_wallaby}. We take the standard deviation of the line-free regions of the spectra at their final chosen resolutions, and remeasure each metric 1000 times with the spectra shifted by a noise array generated by a Gaussian of this width. The final uncertainties are taken as the standard deviation of these measurements. 

There are also a number of contributors to the systematic uncertainty that arises from these methodologies. One source originates from the choice of central velocity used in each of the metrics. As these centres are generally some fractional distance inside a channel, we must interpolate our spectral bins such that the nearest half channel increment aligns with the centre. We estimate the uncertainty in this method in a similar fashion as the statistical contribution, taking 1000 measurements with central velocities drawn from a Gaussian with a mean equal to the true centre and a width of one channel. The standard deviation of these results are taken as our systematic uncertainty. Thus, in our results, we present the value \PM\ $\delta$(stat.) \PM\ $\delta$(sys.)

The \texttt{3DACs} code does not output any uncertainties directly. \citet{Deg2023} claims that in their testing, the uncertainty associated with noise variation rarely exceeds 0.02, and almost certainly is dominated by systematics. We also note that the authors highlight imprecisions in the \texttt{3DACs} results when the size of the \hi\ disk is smaller than four resolution elements (beams). Unfortunately, only three of our targets (FRB 20190425A, FRB 20200723B, and FRB 20211127I) satisfy this requirement. However, we still wish to investigate how useful these metrics are on real datasets across different instruments and resolutions. Thus, we choose to incorporate our own uncertainty associated with the choice of rotation centre --- which is the quantity that introduces the most error due to low spatial resolution --- by again retaking 1000 measurements with centres perturbed by a 3D Gaussian, with $\sigma_\text{x}$, $\sigma_\text{y}$, and $\sigma_{\nu}$ all equal to 1. This is relatively straightforward as \texttt{3DACs} already incorporates interpolation methods to allow for fractional rotation centres. While a Gaussian of this width likely overestimates the uncertainty, it is simple in design and clearly accounts for the effect of spatial and spectral resolution. Furthermore, as we do not present a statistical contribution, we are satisfied with the larger systematic estimate. 

There are also uncertainties that arise due to the effects of galaxy inclination and viewing angle. While \citet{Deg2020} clearly shows that both have notable impacts on these measurements, we do not incorporate such considerations here as their contributions are nontrivial to model and likely require detailed kinematic modelling to accurately describe.

\section{Target Analysis}\label{sec:individuals}

Here we discuss the asymmetry results for each FRB host galaxy, highlighting interesting cases. Table~\ref{tab:asymmetry} presents the quantified asymmetry metrics for each target; at the bottom, we include measurements for the known disturbed host of FRB 20171020A \citep{Lee-waddell23} to compare against.

\begin{table*}
\caption{Asymmetry measurements derived from \hi\ profiles and moment maps: flux asymmetry (A\textsubscript{flux}), spectral asymmetry (A\textsubscript{spec}), one-dimensional morphological asymmetry (A\textsubscript{1D}), two-dimensional map asymmetry (A\textsubscript{2D}), and three-dimensional asymmetry (A\textsubscript{3D}). For the first three columns, the first uncertainty represents the contribution from noise, and the second represents the contribution from resolution. For A\textsubscript{2D} and \AtD, the uncertainties represent the choice of central rotation pixel/voxel.}
\label{tab:asymmetry}
\begin{threeparttable}
\begin{tabular}{lcccccc}
\toprule
\headrow FRB Name & Observation & A\textsubscript{flux} & A\textsubscript{spec} & A\textsubscript{1D} & A\textsubscript{2D}\tnote{$\dagger$} & A\textsubscript{3D}\tnote{$\dagger$} \\
\midrule

FRB 20181220A & GMRT & 1.504 \PM\ 0.649 \PM\ 0.528 & 0.354 \PM\ 0.170 \PM\ 0.168 & 0.161 \PM\ 0.098 \PM\ 0.078 & 0.279 \PM\ 0.086 & 0.463 \PM\ 0.132\\
& FAST & 1.055 \PM\ 0.052 \PM\ 0.046 & 0.217 \PM\ 0.046 \PM\ 0.023 & 0.090 \PM\ 0.032 \PM\ 0.015 & N/A & N/A \\
\midrule
FRB 20181223C & GMRT & 1.146 \PM\ 0.271 \PM\ 0.232 & 0.395 \PM\ 0.158 \PM\ 0.127 & 0.196 \PM\ 0.030 \PM\ 0.067 & 0.259 \PM\ 0.055 & 0.252 \PM\ 0.097\\
& FAST & 1.084 \PM\ 0.044 \PM\ 0.073 & 0.229 \PM\ 0.036 \PM\ 0.102 & 0.112 \PM\ 0.022 \PM\ 0.057 &  N/A & N/A \\
\midrule
FRB 20190418A & GMRT & 1.302 \PM\ 0.375 \PM\ 0.397 & 0.488 \PM\ 0.177 \PM\ 0.168 & 0.224 \PM\ 0.035 \PM\ 0.099 & 0.208 \PM\ 0.041 & 0.232 \PM\ 0.114 \\
\midrule
FRB 20190425A\tnote{$*$} & GMRT & 1.086 \PM\ 0.081 \PM\ 0.062  & 0.175 \PM\ 0.067 \PM\ 0.109 & 0.109 \PM\ 0.003 \PM\ 0.080 & 0.099 \PM\ 0.028 & 0.140 \PM\ 0.062 \\
& FAST & 1.054 \PM\ 0.029 \PM\ 0.027 & 0.171 \PM\ 0.022 \PM\ 0.028 & 0.070 \PM\ 0.016 \PM\ 0.029 & N/A & N/A \\
\midrule

FRB 20200723B\tnote{$*$} & MeerKAT & 1.198 \PM\ 0.058 \PM\ 0.137 & 0.111 \PM\ 0.037 \PM\ 0.142 & 0.058 \PM\ 0.023 \PM\ 0.091 & 0.084 \PM\ 0.037 & 0.210 \PM\ 0.166 \\
\midrule
FRB 20210405I & MeerKAT & 1.135 \PM\ 0.148 \PM\ 0.069 & 0.300 \PM 0.099 \PM\ 0.076 & 0.170 \PM\ 0.026 \PM\ 0.042 & 0.074 \PM\ 0.066 & - \\
\midrule
FRB 20211127I\tnote{$*$} & MeerKAT & 1.034 \PM\ 0.060 \PM\ 0.061 & 0.105 \PM\ 0.056 \PM\ 0.097 & 0.043 \PM\ 0.037 \PM\ 0.060 & 0.091 \PM\ 0.026 & 0.116 \PM\ 0.042 \\
\midrule
FRB 20211212A & MeerKAT & 1.244 \PM\ 0.563 \PM\ 0.299 & 0.371 \PM\ 0.222 \PM\ 0.143 & 0.140 \PM\ 0.048 \PM\ 0.065 & 0.126 \PM\ 0.055 & - \\
\midrule
FRB 20231229A & Arecibo & 1.360 \PM\ 0.079 \PM\ 0.125 & 0.323 \PM\ 0.051 \PM\ 0.042 & 0.140 \PM\ 0.027 \PM\ 0.026 & N/A & N/A \\
\midrule 
FRB 20231230A & MeerKAT & 2.394 \PM\ 1.252 \PM\ 1.161 & 0.311 \PM\ 0.225 \PM\ 0.158 & 0.130 \PM\ 0.046 \PM\ 0.104 & - & - \\
\midrule 
\red{FRB 20240210A}\tnote{$*$} & MeerKAT & 1.048 \PM\ 0.057 \PM\ 0.035 & 0.197 \PM\ 0.053 \PM\ 0.055 & 0.077 \PM\ 0.016 \PM\ 0.035 & 0.134 \PM\ 0.063 & 0.166 \PM\ 0.048 \\
\midrule 
\red{FRB 20240312D}\tnote{$*$} & MeerKAT & 1.161 \PM\ 0.087 \PM\ 0.141 & 0.145 \PM\ 0.054 \PM\ 0.097 & 0.072 \PM\ 0.033 \PM\ 0.058 &  0.133 \PM\ 0.017 & 0.199 \PM\ 0.059 \\
\midrule 
\red{FRB 20240615B} & MeerKAT & 1.208 \PM\ 0.226 \PM\ 0.149 & 0.379 \PM\ 0.123 \PM\ 0.037 & 0.153 \PM\ 0.088 \PM\ 0.024 & 0.053 \PM\ 0.050 & - \\
\midrule 
FRB 20250316A & Effelsberg &1.659 \PM\ 0.187 \PM\ 0.192 & 0.406 \PM\ 0.081 \PM\ 0.015 & 0.180 \PM\ 0.049 \PM\ 0.008 & N/A & N/A \\
\bottomrule
\bottomrule
FRB 20171020A\tnote{$*$} & ATCA & 1.320 \PM\ 0.037 \PM\ 0.184 & 0.319 \PM\ 0.027 \PM\ 0.076 & 0.1633 \PM\ 0.013 \PM\ 0.036 & 0.123 \PM\ 0.016 & 0.237 \PM\ 0.021\\
\bottomrule
\end{tabular}
\begin{tablenotes}[hang]
\item[$\dagger$]Dashed lines indicate measurements with no statistically significant asymmetry (-1 result returned by \texttt{3DACs}).
\item[$*$]Targets with resolutions / SNRs great enough for 2D/3D measurements to be considered robust according to \citet{Deg2023}.
\end{tablenotes}
\end{threeparttable}
\end{table*}

\subsection{FRB 20181220A}

The host of FRB 20181220A is a highly inclined edge-on spiral which shows moderate \hi\ emission in both the GMRT and FAST data. By eye, the higher SNR FAST spectrum appears largely symmetric, which is reflected in its low asymmetry metrics. In contrast, the GMRT spectrum exhibits significantly greater asymmetry, though the deviation is consistent within the noise level. Furthermore, the intensity map shows a peak line flux away from the optical core of the galaxy. This causes the $A_\text{2D}$ and \AtD\ values to inflate to the highest in this sample. However, the area of the GMRT detection with respect to the beam size is right in the grey area of where \citet{Deg2023} identifies that 3D asymmetry modelling becomes less robust. Thus, we are hesitant to place too much weight on the GMRT result, especially considering the spectrum deviates from the FAST spectrum. The GMRT data does reveal that the galaxy lies in an \hi\ rich group, with at least two other nearby detections in the field, the nearest of which lies 0.55 Mpc away.

\subsection{FRB 20181223C}

The spiral host of FRB 20181223C clearly exhibits a disturbed \hi\ distribution in the GMRT intensity map. Once again, the GMRT and FAST spectra do not fully match, but are consistent within the noise when accounting for the much lower SNR of the GMRT detection.  

\begin{figure*}[t!]
    \centering
    \includegraphics[width=\linewidth]{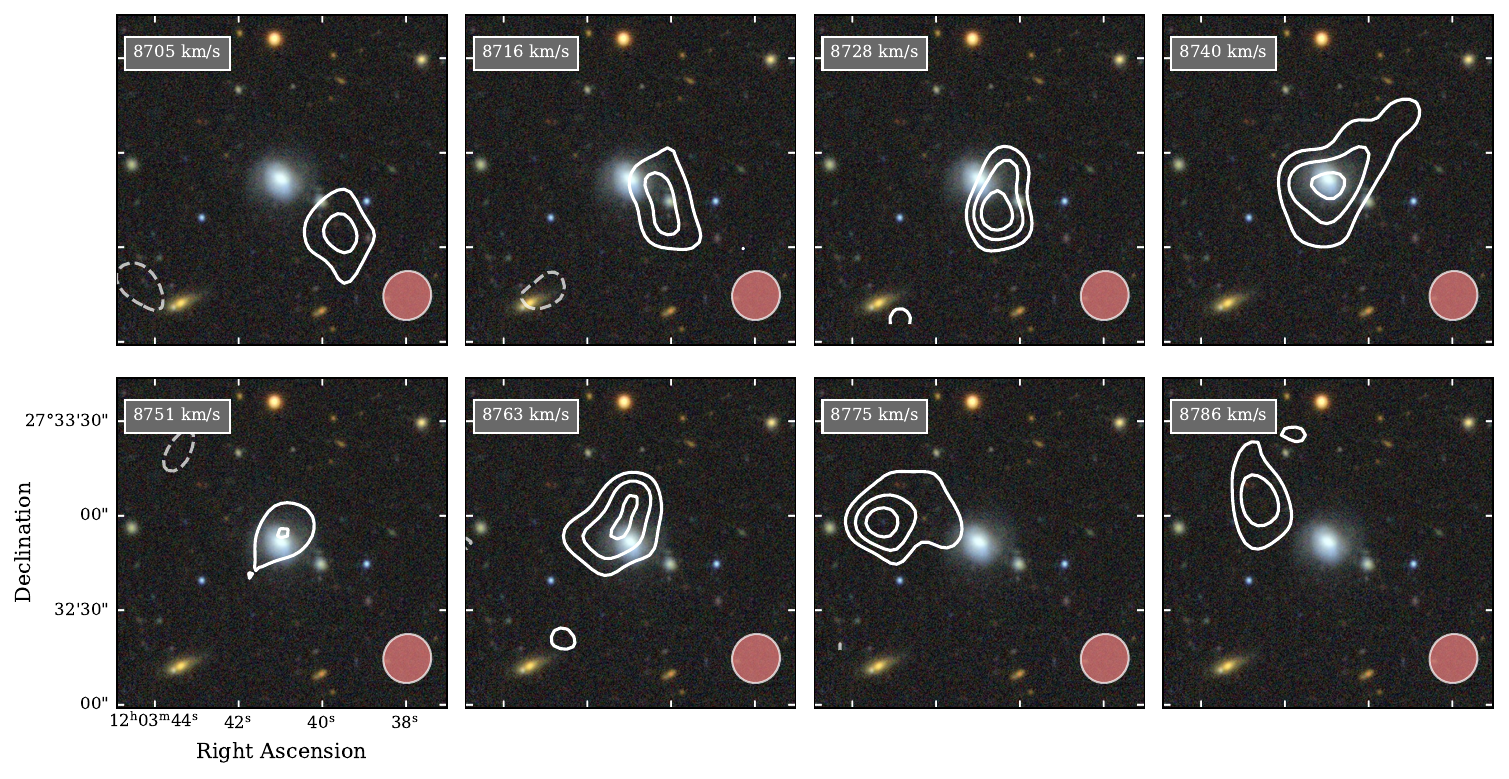}
    \caption{GMRT channel maps of the host of FRB 20181223C, overlaid on DECam imaging. Contours are at 1.0, 1.5, 2.0 mJy beam$^{-1}$, with dashed contours signifying negative counterparts.}
    \label{fig:181223_channelmaps}
\end{figure*}

The moment 0 and 1 maps show a clear extension to the west, captured primarily in a single channel due to the low spectral resolution of the cube. To highlight this, we produce individual channel maps in Figure~\ref{fig:181223_channelmaps}, with the channel at 8775 \kms revealing strong \hi\ emission centred nearly 30" from the core of the galaxy. This is the primary cause of the higher $A_\text{2D}$ and \AtD\ values exhibited by this target. We also note the interesting extension in the channel map at 8740 \kms, which is too faint to affect the total intensity map. 

This host is clearly significantly disturbed, though the origin of this morphology is difficult to ascertain. There are a number of galaxies in the visual field, including one very nearby to the south east, and another further away to the west. 
However, \citet{chimefrbs} investigated these as potential hosts of the FRB, and found incompatible redshifts greater than 0.25 for both. Thus, there is no clear candidate that could be causing a disturbance through tidal means.  

%very low potential nearby hi detection

\subsection{FRB 20190418A}

With the third most distant redshift in this sample, the massive host of FRB 20190418A exhibits a weak \hi\ signal in the GMRT data. As such, its spectral features are largely dominated by the noise. However, the moment maps of Fig.~\ref{fig:momentmaps} reveal an asymmetry in the extent of the galaxy either side of its core which gets picked up by SoFiA-2 even with a higher S+C threshold of 5. Furthermore, the asymmetry metrics all suggest disturbance; \Aflux\ exceeds our threshold, \Aspec\ is the highest in this sample, and even \AoD\ --- which is the most resilient to noise --- is \red{also} the highest in this sample. Thus, we consider this target disturbed, though deeper observations would be useful to confirm the nature of this. 

% The velocity map suggests that the south west region of the galaxy is far less settled than the blueshifted north east, though the significance of this is difficult to ascertain. It is worth noting that this host is closest in Figure~\ref{fig:hi_properties} (c) to the clearly disturbed FRB host presented in \citet{Kaur2022}, lying well above the median \hi\ mass for its sSFR. This may be an indicator of disturbance, though given the SNR and beam size of this observation, it is certainly possible that deeper observations would reveal a smooth disk profile similar to FRB 20190425. As such, we consider this target potentially disturbed, but would require deeper observations to confirm this. 

%nearby detections dont  feel real 

\subsection{FRB 20190425}

The host of FRB 20190425A appears to be largely undisturbed. The FAST and GMRT spectra are symmetric, and the moment maps are typical for a spiral of this inclination. This is reflected in the low asymmetry metrics across the board, all of which are robust due to the spectral and spatial resolution of the cube. This galaxy also appears isolated, with no nearby \hi\ detections observed.

%clearly isolated

\subsection{FRB 20200723B}

\begin{figure*}[t!]
    \centering
    \includegraphics[width=\linewidth]{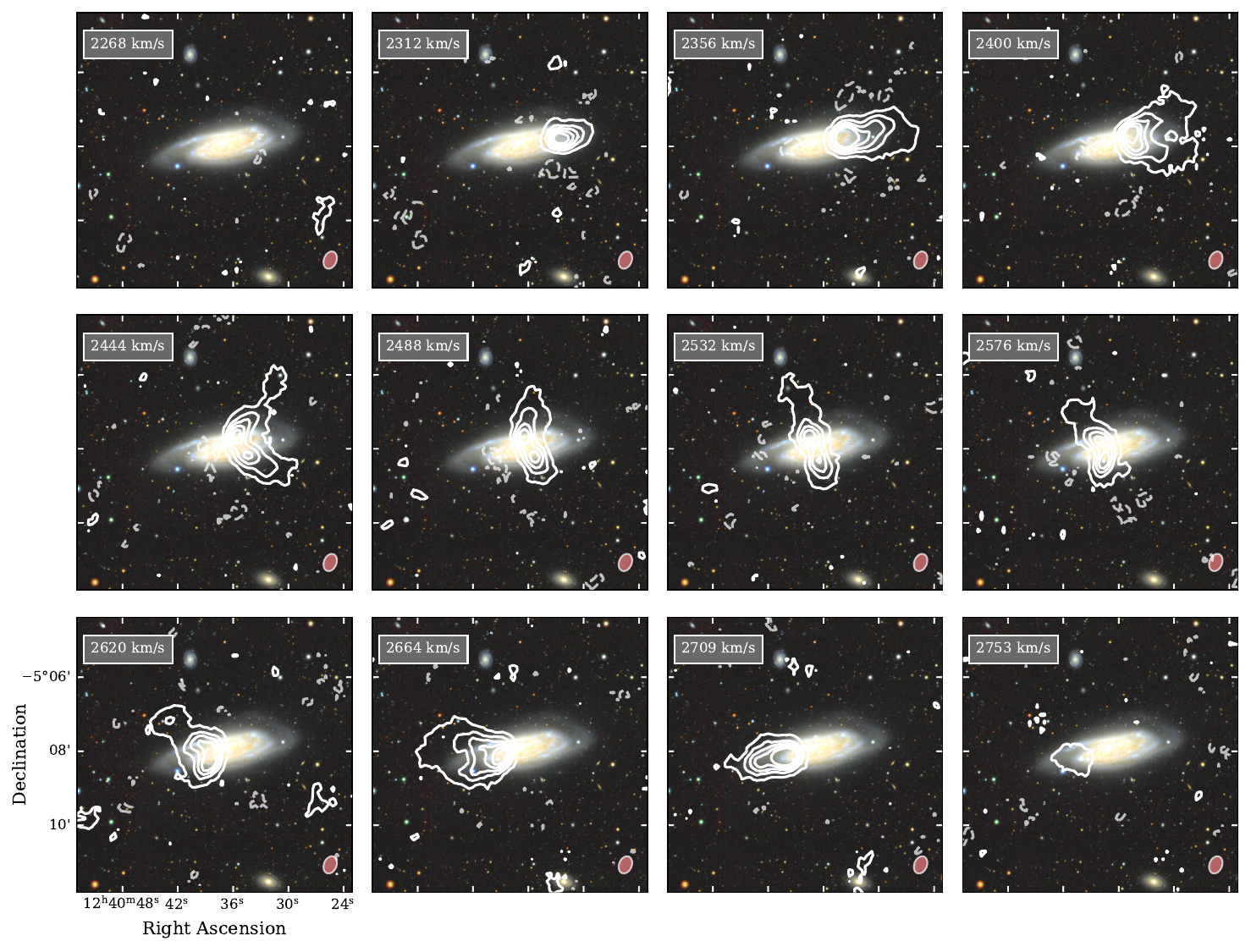}
    \caption{MeerKAT individual channel maps of the host of FRB 20200723B, overlaid on DECam imaging. Contours are at 1, 3, 5, 7, and 9 mJy beam$^{-1}$, with dashed contours signifying negative counterparts.}
    \label{fig:FRB20200723b_channelmaps}
\end{figure*}

The host of FRB 20200723B again 
exhibits mostly typical \hi\ emission, with a symmetrical spectrum and emission aligned with the plane of the optical disk. However, the northern edge of the galaxy appears a lot less smooth than the rest; investigating the individual channel maps reveals faint elongation in at least five channels out to approximately 20 kpc above the stellar disk, as shown in Fig.~\ref{fig:FRB20200723b_channelmaps}. To confirm the nature of this elongation, we employ \texttt{3DBarolo} to fit a smooth rotation disk profile, and compare with the data (see Fig.~\ref{fig:barolo} in Appendix), which verifies this emission is not standard for a settled galaxy. The flux level of these features is near the noise level, and thus they are not contributing significantly to the asymmetry metrics, though the \AtD\ result is not far from that of the disturbed host of FRB 20171020A. We also note that the DECam photometry hints at a potential perturbation to the stellar distribution on the north east side, with the spiral arm extending in a slightly unexpected direction to the north.

% To fully highlight the disturbance, we employ \texttt{3DBarolo} to fit a smooth rotation disk profile, and compare with the data,

As the disturbance is spread across the entire northern face, and the velocity map shows no evidence of the more uniform bulk motion expected from a wind-driven outflow, this feature is most likely caused by a recent or ongoing minor tidal interaction. \red{There are a number of nearby extended sources in the DECam imaging, though the only \hi\ detection visible in MeerKAT arises from a large spiral 1.6 Mpc to the south.} The tip of the outer contour in the 2532 \kms channel map, which shows the most extreme elongation, appears to overlap with a faint extended source, \red{though this is likely a coincidence}. Unfortunately, with no further redshift information on the surrounding objects, it is not possible to definitively \red{characterise the system}. Regardless, these \hi\ observations reveal that NGC 4602 is another disturbed FRB host galaxy.

%rich HI group 10'

% Such a feature may result from a number of circumstances, which we consider here. NGC 4602 is a moderately star forming galaxy; from its bright RC emission, we derive a SFR of 3.21 \msun yr$^{-1}$, which broadly aligns with the H$\alpha$ derived value of 4.5 \msun yr$^{-1}$ \citep{Kennicutt83}. It is thus capable of producing supernovae driven outflows, which are known to rise several tens of kpcs above the optical disk. However, the channel maps and photometry show that the disturbance is spread across the entire northern face, rather than just a small region that may arise from a single outflow near a denser region of star formation. The velocity map does not suggest that the gas is flowing away from the disk in a uniform manner as would be expected for an outflow, and instead shows that the gas is rotating with the rest of the galaxy, just at an extended distance from the plane.

% This suggests that a recent minor interaction with a counterpart may have gravitationally pulled the outer gas of the northern edge. There are several extended sources near NGC4602, including the face on spiral 2MASX J12404071-0505321, several fainter blobs that can be seen in the DECAm imaging, and a few RC sources with no clear optical association. Unfortunately, with no further redshift information on the surrounding objects, it is not possible to definitively confirm this scenario. However, we postulate that either a  dwarf companion galaxy, or perhaps even the spiral, is in the process of tidally disrupting NGC 4602. 

\subsection{FRB 20210405I}

The host of FRB 20210405I appears almostly perfectly symmetrical in the MeerKAT data. The \texttt{3DACs} code returned an \AtD value of -1, signifying no statistically significant asymmetry, and thus we conclude this host is undisturbed. According to the SoFiA-2 source finder, this galaxy is also isolated in \hi.

%isolated galaxy

% The host of FRB 20210405I exhibits a pronounced \hi\ distribution which reaches far beyond its optical extent. This was relatively faint in the full 64K resolution data, and thus we rebinned to 32K, 16K, 8K, and 4K resolutions. Investigating each of the datacubes revealed a potential faint feature 

% , and displays signs of tidal interactions. The galaxy is situated in the Zone of Avoidance (ZoA), the region of the sky most significantly impacted by extinction from the Milky Way. As a result, the optical emission - which manifests as an extended blob approxmiately 12 kpc across - is likely stunted as seen from Earth. However, the \hi\ cloud extends out to a radius of nearly 100 kpc, which likely implies that the galaxy is indeed massive. 

% In the initial cleaned data cube, we noticed 

% The northern end of the \hi\ cloud exhibits an extended feature, which gets picked up by the SoFiA-2 source finding 

% is extended in a manner reminiscent of tidal interactions 

% \citet{Driessen2024} estimated its stellar mass using 2MASS K band imaging and the mass-luminosity relation for galaxies, finding a value of log(\mstar/\msun) = 11.25. They also found an upper limit on its \hi\ mass from archival MeerKAT observations

\subsection{FRB 20211127I}

The host of FRB 20211127I was previously analysed in \citet{Glowacki_wallaby} using low resolution ASKAP data, and was prior to this study the only non-interacting FRB host observed in \hi. The detection appeared symmetric in spatial and spectral dimensions, but the authors concluded that higher resolution data would be required to fully confirm this case. 

The new MeerKAT observation does indeed reveal a mostly symmetrical \hi\ distribution, both in the spectra and intensity map. The 1D asymmetry measurements are the lowest in this study, and while \AtD value returned by \texttt{3DACs} is not -1, it is low. This is perhaps due to the slight elongation in the north east corner of the intensity map, but neither the channel nor velocity maps indicate anything definitive. Thus, we conclude that this host galaxy is undisturbed. Interestingly, it is worth noting that unlike the previous two undisturbed hosts, this galaxy is in a rich \hi\ group. Three nearby galaxies are detected in this deeper MeerKAT observation, none of which were present in the WALLABY Pilot Survey II observation of this field \citep{Glowacki_wallaby}. The nearest neighbour lies roughly 2.4 Mpc away.

\subsection{FRB 20211212A}

The faint emission from this high redshift host appears to be symmetrical, though neither the spectral nor spatial resolution is great enough to properly describe it. We conclude that it is mostly likely an undisturbed galaxy, but deeper observations would be needed to confirm this. This host has two faint \hi\ neighbours that each lie approximately 2 Mpc away.

\subsection{FRB 20231229A}

The host of FRB 20231229A is a face-on spiral that appears to have a significant disturbance to one of its spiral arms. The Arecibo spectrum shows a moderately asymmetric profile, resulting in an \Aflux\ value above our threshold for asymmetry. We consider it likely that this target is disturbed, but interferometric observations are required to fully reveal the nature of this.  

\subsection{FRB 20231230A}

As this target was detected in an off-centred observation, the noise level dominates its \hi\ spectrum. It appears slightly asymmetric, but with such low resolution data, is not possible to accurately quantify, as shown by the enormous uncertainties in the asymmetry metrics. Targeted observations are required to ascertain the true nature of this host.

%neighbour 14' away

% The host of FRB 20231230A exhibits \hi\ emission that is only just visible in the off-centred MeerKAT observation we acquired. Due to its location, we did not do a full \texttt{katbeam} primary beam correction at the datacube level, instead simply applying a correction factor to the extracted spectrum given by $S_{\text{corrected}}(\nu)=S_{\text{raw}}(\nu) / \exp(-r^2/2\sigma^2)$ where $r$ is the angular distance from the centre of the beam and $\sigma$ is the standard deviation of the assumed MeerKAT Gaussian beam profile.

% Due to its low SNR and resolution, it is difficult to fully characterise the \hi\ distribution, but it appears to be somewhat asymmetric in the spectral domain. However, this could easily be a product of the noise level. Legacy DECam imaging does not suggest any disturbance, and the moment maps are limited in the information they carry. With an estimated \hi\ mass of 6.70 \tenpow{9} \msun, this galaxy is certainly a prime candidate for future targeted observations. 

%A bright face-on spiral, the host has a 2MASS $K$ band magnitude of 13.122; assuming a simple mass-to-light relation of $M_{\star}/L_K\sim0.6$ \citep[in solar units, a decent estimation for low redshift spirals,][]{McGaugh2014}, this corresponds to an approximate stellar mass of 1.15 \tenpow{10} \msun. However, the MeerKAT observations suggest that the both the \hi\ mass and SFR of the host are low, at 

\subsection{FRB 20240201A}

The host of FRB 20240201A is an edge-on spiral galaxy in an apparent galaxy pair. Spectroscopic observations from the Sloan Digital Sky Survey (SDSS) reveal the two share similar redshifts; 0.04273 and 0.04314 for the host and neighbour respectively. The moment maps clearly show an overlap in the clouds surrounding the two galaxies. The data displayed was processed with a robust weighting of 0.5; a weighting of 2 reveals the extent of the combined gas cloud reaches far beyond the combined optical region, whilst a weighting of 0 is unable to pick up the emission of the FRB host. 

\red{The spectra presented in Fig.~\ref{fig:spectra} show the superposition of both galaxies' emission. The host contributes partially to the first peak but is dominated by the neighbour; examining the resolved MeerKAT data reveals the neighbour's rotating arms form the bulk of the outer peaks and its notably gas rich central bulge forms the third peak in the centre.} The slightly blueshifted nature of the \hi\ emission originating from the position of the FRB host, seen in the MeerKAT moment 1 map, confirms that it is not simply emission from the edges of the neighbour's \hi\ cloud. 

Inspecting the SoFiA-2 mask confirms that there is a smooth emitting connection in both physical and spectral space between the two galaxies. From this, it is clear that the pair are tidally interacting, and likely have been for some time. A good amount of emission from the neighbour is arising from directly behind the host, and thus it is not possible to cleanly extract an \hi\ spectra for it alone. As such, we calculate global properties for the pair as a whole, and do not attempt to measure the asymmetry metrics. This interaction observed in \hi\ correlates with the apparent distortion of the stellar disk in the south west of the neighbouring galaxy towards the FRB host, as seen in the DECam imaging. The velocity dispersion map reveals heightened turbulence in the region between the two, highlighting the impact of their gravitational influences. 
% While this system does lie near several other \hi\ detections, their 3D projections imply they all lie further than 10 Mpc away. 

% The combined \hi\ region totals to a mass of 9.41 $\times$ 10$^{9}$ \msun from the MeerKAT data, and 7.44\tenpow{9} \msun according to the FAST data. We present the MeerKAT observation numbers in Table~\ref{tab:radioproperties} due to its resolved nature and more robust processing.

% Both the host and neighbour are coincident with radio continuum sources: an unresolved 0.42 mJy source and a resolved 4.71 mJy source respectively. Using these, we estimate the global SFR of the host and neighbour to be 0.91 \msun yr$^{-1}$ and 6.65 \msun yr$^{-1}$ respectively. While no explicit SFR has been derived in the literature from other regimes, the SDSS spectroscopy includes clear H$\alpha$ emission lines from which an estimate can be made...

\subsection{FRB 20240210A}

\red{The spiral host of FRB 20240210A lies only 1.5'
from of a smaller neighbour to the south west, and both exhibit clear \hi\ emission. When imaging with a Briggs weighting with a robust value of 2, SoFiA-2 considers the galaxies to be connected. However, the significance of any potential tidal bridge is extremely marginal; even when rebinned from a spectral resolution of 26 kHz to 104 and 208 kHz resolutions, the link never rises above 2$\sigma$, and is completely absent when imaging with a robust weighting of 0.5. Furthermore, individual channel maps do not suggest any consistent extension of the host galaxy.}

\red{Given the neighbour's archival redshift of 0.02360 from the 2dF Galaxy Redshift Survey \citep{2dFGRS}, the angular separation deprojects into a physical distance of roughly 1 Mpc. The spectral profile of the host is symmetric, and none of the asymmetry metrics hint at intrinsic disruption. As such, we do not consider this host to be tidally interacting with the neighbour, and thus is undisturbed.}  

\subsection{FRB 20240312D}

\begin{figure*}[t!]
    \centering
    \includegraphics[width=\linewidth]{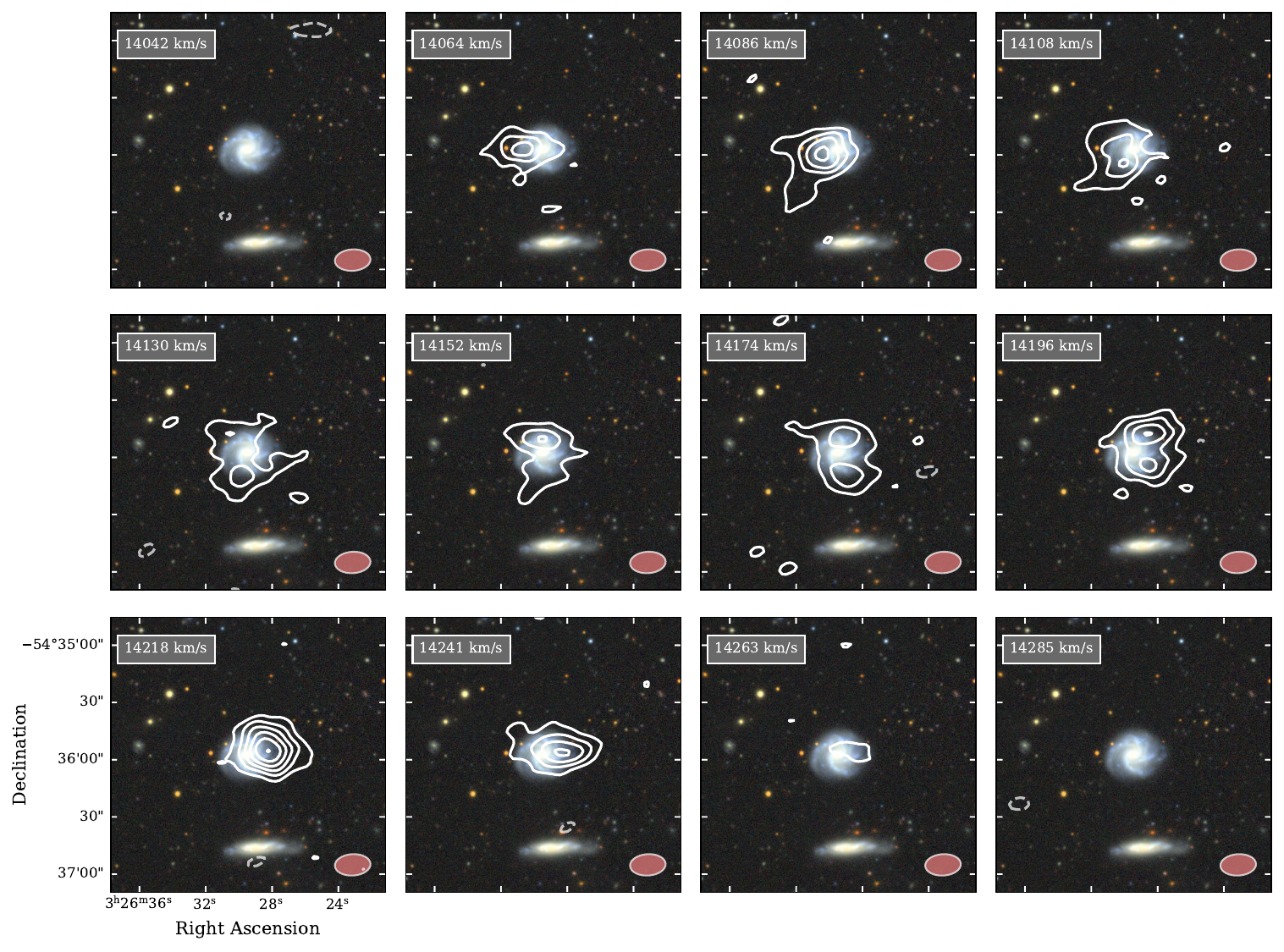}
    \caption{MeerKAT individual channel maps of the host of FRB 20240312D, overlaid on DECam imaging. Contours are at 0.4, 0.8, 1.2, 1.6, 2.0, and 2.4 mJy beam$^{-1}$, with dashed contours signifying negative counterparts.}
    \label{fig:240312_channelmaps}
\end{figure*}

\red{The host of FRB 20240312D shows an unusual gas distribution in its intensity map, with the emission extending further to the south than to the north. Channel maps, presented in Fig.~\ref{fig:240312_channelmaps}, confirm this asymmetry, as the southern extension is present across at least a third of the velocity profile. However, these features are less significant than in previous hosts: they do not reach the high contour levels seen in the host of FRB 20181223C, nor are they highly resolved by the beam as in the host of FRB 20200723B, making the evidence less conclusive. This is further reflected in the host's \AtD\ value of 0.199, which sits above every undisturbed host and below every disturbed host.}

\red{This galaxy resides in a highly rich \hi\ group containing at least eight nearby \hi\ galaxies, a number of which are within 1 Mpc. Cross-matching these detections with DECam imaging reveals that, apart from one case, they are all small, faint dwarf galaxies. The larger edge-on spiral visible to the south in Fig.~\ref{fig:240312_channelmaps} appears close in projection, but the central frequency of its strong \hi\ line places it nearly 10 Mpc in the background. Consequently, the nearest companion to the FRB host lies roughly 600 kpc to the east. Given the host exists in such a dense environment, and the observed gas extension does not point toward the nearest neighbour, it is more likely that the asymmetry is simply a lingering imprint of an earlier gravitational perturbation caused during the formation of the group than a recent tidal interaction. In these cases, peripheral gas can take considerable time to resettle into the disk. We therefore regard this host as a middle-ground case and, in Section~\ref{sec:discussion}, examine it under both classifications when interpreting the results.}

\subsection{FRB 20240615B}

\red{The \hi\ emission arising from this distant host appears largely symmetrical. The line is narrow, but the double-horned profile is still resolved, with the peaks reaching similar intensities. Neither its moment maps nor its asymmetry metrics suggest any disturbance, and thus we conclude this host is settled. It is worth noting that this galaxy does lie in a relatively \hi\ rich group, with at least four galaxies within 10 Mpc, the nearest positioned roughly 900 kpc away.}  

\subsection{FRB 20250316A}

NGC 4141 is a slightly inclined spiral that shows a highly asymmetric line profile in the archival Effelsberg spectrum. In particular, its \Aflux\ parameter is the highest robust measurement in this sample. PanSTARRS imaging reveals that a highly elongated uncatalogued galaxy lies very nearby; it appears very similar in colour profile to NGC 4141, and thus may well have a similar redshift. If this were to be the case, the two would almost certainly be interacting, which would likely be the cause of the profile asymmetry. Therefore, this region is a prime target for future interferometric observations to resolve its spatial \hi\ distribution.

\section{Discussion}\label{sec:discussion}

\subsection{Are FRB Hosts Disturbed?}\label{subsec:disturbed}

From this analysis, we find that six of these FRB host galaxies are disturbed: FRB 20181223C, FRB 20190418A, FRB 20200723B, FRB 20231229A, FRB 20240201A, and FRB 20250316A. \red{Another six are clearly undisturbed: FRB 20190425A, FRB 20210405I, FRB 20211127I, FRB 20211212A, FRB 202404210A, and FRB 20240615B. One is borderline (FRB 20240312D), and the remaining three require deeper/targeted observations to draw firm conclusions.} With respect to the total population of FRB hosts now observed in \hi\, the confirmed ratio of disturbed to undisturbed galaxies is \red{11:6}. 

Comparing this ratio to background populations is somewhat nebulous. While in Section \ref{subsec:asym_thresh} we quantify the thresholds for disturbance with respect to the asymmetry metrics, we have not relied solely on these values as indicators of disruption. In fact, at least two of the hosts we consider disturbed here may not have been considered so in large volume surveys. For example, determining the disrupted nature of the host of FRB 20200723B required in-depth analysis of individual channel maps at low flux thresholds; none of the asymmetry metrics would otherwise suggest this nature. It is likely that digging deep enough would reveal that the fraction of nearby galaxies with some low level disruption similar to this is much higher than the fractions reported from asymmetry quantification methods. 

\begin{figure}[t!]
    \centering
    \includegraphics[width=\linewidth]{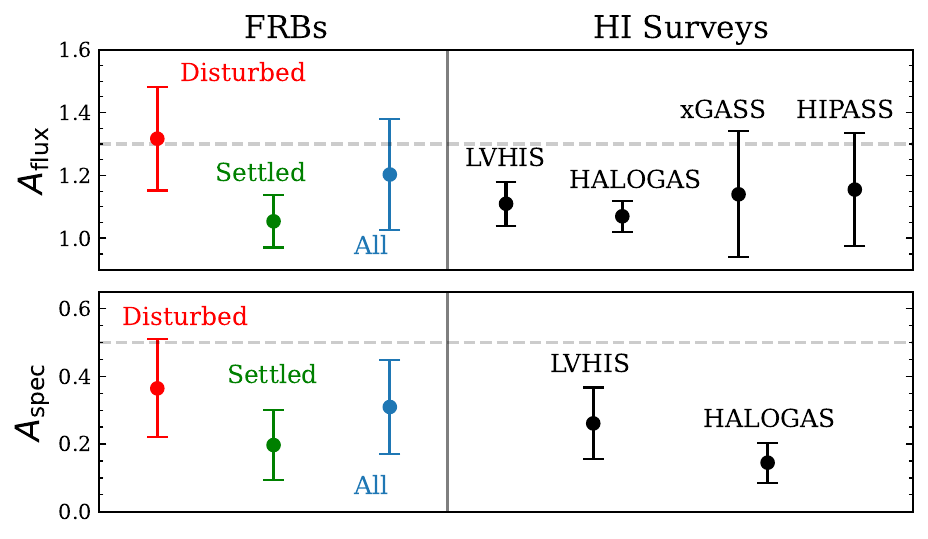}
    \caption{\Aflux\ and \Aspec\ values of FRB hosts compared to background populations presented in \citet{Reynolds2020LVHIS,Watts2020}. LVHIS and HALOGAS sample low-density environments, whereas xGASS is more representative of the local galaxy population. The xGASS result is an approximation; we generate the results by extracting the cumulative distribution of Fig. 2 from \citet{Watts2020}. The two dashed lines indicate our chosen thresholds for asymmetry, outlined in Sec.~\ref{subsec:asym_thresh}.}
    \label{fig:asym_comp}
\end{figure}

Therefore, the observation that $\sim$65\%\ of FRB hosts are disturbed cannot be directly contrasted to the 10-40\%\ results in the literature. In Figure \ref{fig:asym_comp}, we do utilise the \Aflux\ and \Aspec\ metrics to compare with the LVHIS and HALOGAS samples as is done in \citet{Michalowski21,Glowacki_hi_frb_whats_your_z}, and also include approximations for the xGASS \citep{xGASS} and HIPASS \citep{Barnes2001} datasets. These are stellar mass and \hi\ mass selected samples that span a wide range of environments and masses, and thus are more representative of the local galaxy population. We estimate their \Aflux\ values by extracting the data from \citet[Fig. 2]{Watts2020} and \citet[Table 5]{Reynolds2020} respectively. We note that robust comparison should be done exclusively with galaxies similar in mass and SFR to FRB hosts; however, both \cite{Watts2020,Reynolds2020LVHIS} find no evolution in \Aflux\ or \Aspec\ as a function of mass, and thus comparing with these samples is satisfactory. Our sample of FRB hosts exhibits slightly higher 1D asymmetry levels than the background galaxy population, though the discrepancy is minimal --- particularly when compared to the more representative background samples.

As discussed in Section \ref{subsec:asym_thresh}, comparing our \AoD, $A_{2D}$, and \AtD\ values to literature is largely non-viable. \citet{Perron-Cormier2025} only includes figures of their \AoD\ and \AtD\ distributions for the $\sim$31\% of galaxies in their sample which had \AtD\ $\geq0$ (i.e.\ not -1). Thus, these results are inherently biased against settled galaxies. 

All this considered, we cannot definitively claim what the disturbed-to-settled ratio for background galaxies is in a way that is meaningful for comparison with our sample. We speculate that it should hover somewhere near 1:1 considering the quantified asymmetry percentages of background galaxies and including some allowance for the more detailed analysis we have used to characterise our sample. Under this assumption, the observed rate of disturbance in FRB hosts remains higher than expected.

To quantitatively assess the significance of this, we can assume Poisson error, and perform a likelihood ratio test. If our null hypothesis is that the ratio should follow 1:1 --- i.e. the Poisson $\lambda$ variable for both the disturbed and undisturbed host galaxy counts is 8.5 in this case --- then a result of \red{11:6 results in a p-value associated with rejecting this hypothesis of 0.222. This is far higher than the standard significance threshold of 0.05; even considering the borderline case of FRB 20240312D as a disturbed host only shifts the p-value to 0.153. As such, we find that the earlier apparent trend suggesting underlying disturbance in FRB host galaxies is no longer statistically significant. While our observations, which reveal that 21 of 22 FRB hosts emit in \hi, further link FRB progenitors with star formation, the data provide no evidence that merger-induced star formation enhancement plays a significant role in their production.}

% Where does that leave the hypothesis regarding FRBs and the disruption in their hosts? Clearly, any connection between their progenitors and an increased star formation rate due to tidal disturbances is not consistent. This does not preclude the fast channel FRB progenitor theory, but, considering the borderline disturbance trend and the varied gas fractions of FRB hosts (see Sec.~\ref{subsec:hi_prop}), it is unclear whether merger-induced star formation is a dominant driving factor in these galaxies as has been postulated in previous studies. 

% We do speculate that the 

% As such, we believe that the trend of asymmetry in FRB hosts is unlikely to be a real phenomenon, and is probably a result of low-number statistics. This does not necessarily argue against the fast channel FRB progenitor theory, but, considering the varied gas fractions of FRB hosts (see Sec.~\ref{subsec:hi_prop}), it is unlikely that merger-induced star formation is a dominant driving factor in these galaxies. It would still be worth following up a number of these targets and the hosts of future low-z FRBs to fully solidify this conclusion and bring the ratio of disturbed to settled FRB hosts to a more natural state. 

 % This result is similar to many of the population studies when looking at the hosts of FRBs; e.g.\ most hosts are associated with star-formation, but a few cases, such as the globular cluster FRB, conflict with this.

\subsection{\hi\ Properties of FRB Hosts}\label{subsec:hi_prop}

\begin{figure}[t!]
    \centering
    \includegraphics[width=0.92\linewidth]{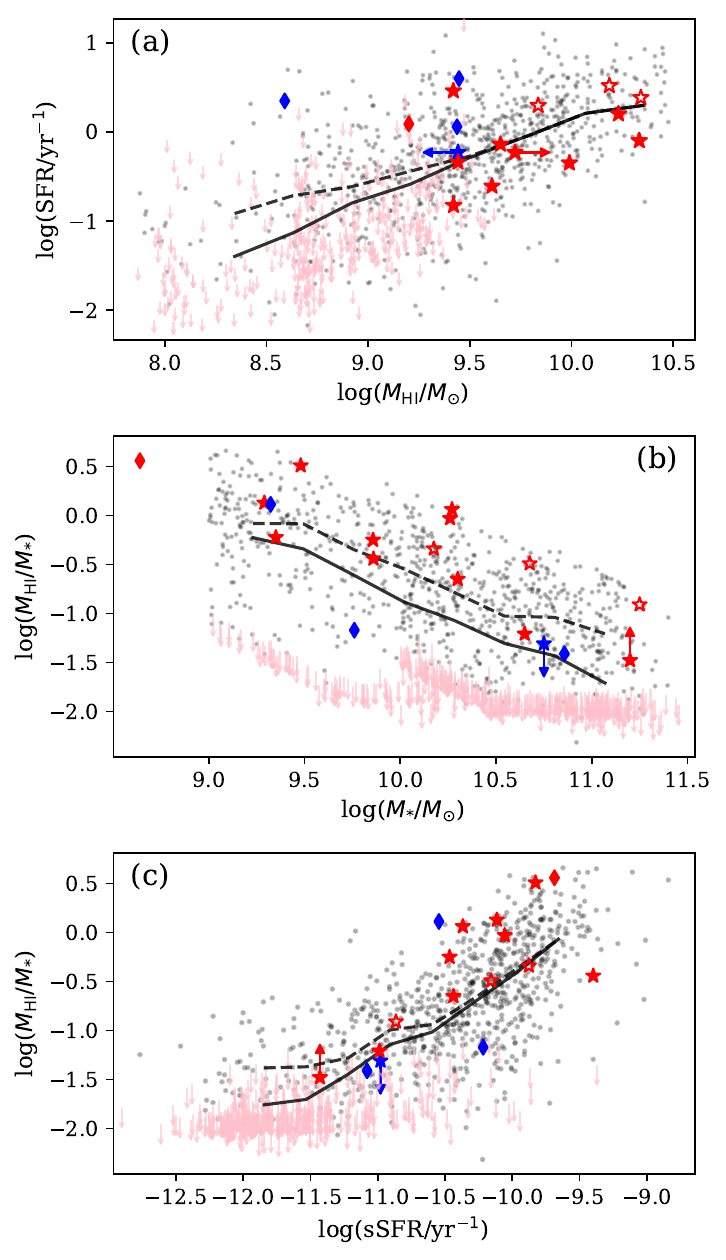}
    \caption{Global \hi\ properties of FRB hosts, separated into non-repeating (red) and repeating (blue) FRBs. Stars indicate our sample, and diamonds indicate the previously published hosts. These are overlaid on the xGASS sample (grey = detected in \hi, pink = non-detections in \hi) \citep{xGASS}. The FRB host stellar masses and SFRs are taken from literature SED modelling values where possible; \red{open stars represent hosts without SED fitting, where we derive $M_*$ from literature $K$-band luminosities and use our radio continuum SFR estimates (see Table~\ref{tab:radioproperties})}. The black lines indicate the median values at eight arbitrary bins along the x-axis for all xGASS galaxies (solid line) and for just \hi\ detected galaxies (dashed line).}
    \label{fig:hi_properties}
\end{figure}

As with most other galaxy population measurements, FRB hosts appear to span a broad range of values with respect to their global \hi\ properties. In Figure~\ref{fig:hi_properties}, we compare our hosts with the xGASS dataset \citep{xGASS} and with previously analysed FRB hosts. Hosts of non-repeaters are shown in red, and hosts of repeaters are shown in blue. All of our new FRBs are non-repeaters, other than FRB 20200223B, whose host is our only non-detection. Measurements of stellar mass and SFR in the xGASS sample are derived from UV and mid-IR observations, tracing the past 100 Myrs of star-formation history (SFH). The values we use for our FRB hosts are taken from SED fitting in the respective localisation papers (see Table~\ref{frbhostoverview}). \red{The hosts of FRB 20210405I, 20240312D, and 20240615B have not had SED fitting, and thus we derive stellar mass approximations using their 2MASS $K$-band magnitudes and the mass-luminosity relation for spiral galaxies \citep{McGaugh2014}, and use their RC-derived SFRs, which should probe similar timescales of SFH to SED fitting; these hosts are shown with unfilled stars in Fig.~\ref{fig:hi_properties}}.

\red{Panel (a) shows that FRB hosts have standard SFRs for their \hi\ masses, but are mostly massive in \hi\ ($M_{\rm HI} > 10^{9.5} M_\odot$) with respect to the background population, which aligns with the \red{stellar mass} trend observed in optical studies \citep{Gordon2023,Sharma2024}. Interestingly, the four repeater hosts appear to lie towards the less massive end of the FRB host \hi\ mass range; \citet{Gordon2023,Sharma2024} also noted a similar pattern in their stellar masses. However, due to the low number of repeaters in their samples (6 and 5), significance tests found no statistical evidence for such a trend. However, \citet{Chen2025} presented a comparison between \mstar\ and SFR(H$\alpha$) in 11 repeater and 33 non-repeater hosts, and found a  statistically significant separation (p-value of 0.0203 when excluding high redshift non-repeaters) between the two populations when considering both properties in tandem.} 

\red{Only one of the repeater hosts in our sample was included in any of the stellar mass studies above, potentially providing further evidence to separate repeater and non-repeater host populations based on mass. However, with such a small number of targets, and given the level of overlap observed in panel (a), this phenomenon is clearly not yet significant. To verify this, we run a simple two sample Kolmogorov-Smirnov (KS) test on the \hi\ masses of our sample, and also run a MANOVA (Multivariate Analysis of Variance) test \citep[as in][]{Chen2025} on the multivariate domain of \mhi\ vs. SFR. In each case, we find $p$-values greater than 0.05 (0.11 and 0.16 respectively). Regardless, future \hi\ studies of FRB hosts should consider probing this relation a focus.}

Panel (b) of Fig.~\ref{fig:hi_properties} shows that the average gas fractions (\mhi/\mstar) of FRB hosts are greater than background galaxies; because all of our hosts are spirals, this is expected. With respect to xGASS galaxies with \hi\ detections (which are much more likely to be late-type galaxies), the FRB hosts are slightly gas-rich, but are spread widely about the median.

Panel (c) displays a more robust relationship between gas fraction and specific SFR (sSFR), used by \citet{Kaur2022,Glowacki_wallaby} to assess if any large gas mass had recently been deposited into the galaxy through a merger. FRB hosts tend to lie above the median for \hi\ detected xGASS galaxies, but the spread again is quite large. The highly interacting host of FRB 20180916B (upper blue diamond) lies further above its respective bin median than any in our new sample; however, the two disturbed hosts presented in \citet{Michalowski21} fall below this median, demonstrating that this relationship is not necessarily a consistent indicator of overall disturbance. 

We see no noticeable trend in any of these three plots when colour-coding by our disturbance assessment (see Fig.~\ref{fig:asym_hi} in Appendix). We also see no significant difference between the disturbance in isolated galaxies vs. those in \hi\ groups.

\subsection{The Baryon Fraction of FRB Hosts}

\begin{figure}[t]
    \centering
    \includegraphics[width=\linewidth]{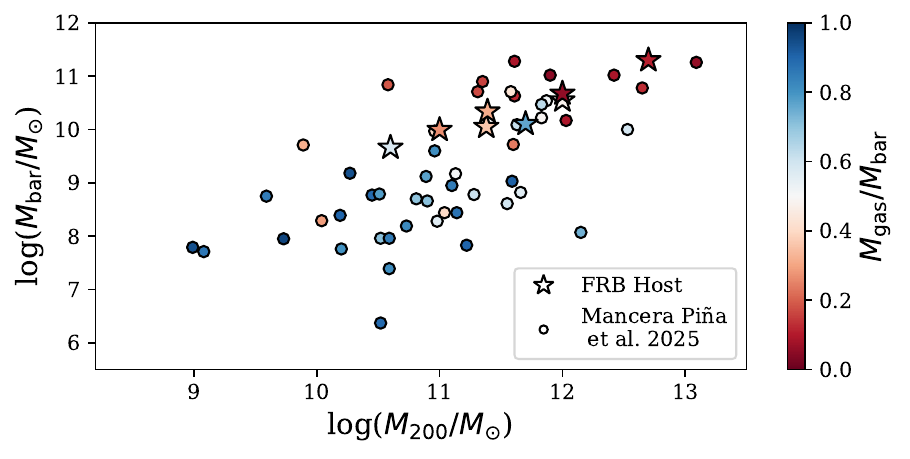}
    \caption{Baryonic vs. virial masses for FRB hosts and the sample of \citet{Pina2025}.}
    \label{fig:m200}
\end{figure}

We briefly consider the baryon fractions, (\mstar+\mhi)/$M_{200}$, of FRB host galaxies (excluding the CGM). We reiterate that our estimates of $M_{200}$ are not particularly robust due to the unconsidered errors in $V_\text{rot}$ arising from inclination uncertainties and instrumental effects. Figure~\ref{fig:m200} compares the galaxy baryonic mass $M_\text{bar}$=\mstar\ + \mhi\ with $M_{200}$ for our selection of hosts and a recent background sample of galaxies from \citet{Pina2025}. It is clear that these FRB hosts are typical when considering this relation, falling generally in line with the background sample in both spatial distribution and gas-baryon fraction, which is indicated by colour. The scatter in this relation is wide, and thus even our rough estimates for $M_{200}$ allow this conclusion to be drawn.

\subsection{Connecting \hi\ Observations with the DM of FRB Hosts}

In addition to using \hi\ to probe the properties of FRB hosts beyond the stellar disk, such observations can be broadly utilised as a proxy for electron column density. Disentangling the host galaxy contribution from the measured dispersion measure (DM) is a pressing issue for cosmological studies using FRBs \citep[for a review, see][]{Glowacki2025}. The free electrons which contribute to the DM arise from the ionised hydrogen (\hii) gas; assuming standard ionisation fractions, \hii\ column densities can be derived from \hi\ column densities. As such, highly resolved \hi\ maps could theoretically be used to better pin down the contribution of the host to the DM.

Local Universe FRBs are particularly useful for this approach. Any uncertainty in the excess DM of an FRB - the remaining contribution when subtracting the Milky Way's disk and halo contributions and the median intergalactic medium contribution as predicted by the Macquart relation \citep{Macquart2020} --- will not be dominated by the scatter on the Macquart relation, and instead will encode information about errors in the estimations of the Milky Way ISM and halo contributions. Recording the \hi\ surface density at the location of an FRB thus may offer clues as to the expected density of free electrons along the line of sight in the host galaxy. Of course, a number of drawbacks --- e.g. the lack of knowledge about the projected distance of the progenitor inside the host, and the quality of the assumptions made regarding ionisation fractions --- limit the efficacy of this approach. However, given the lack of alternative prior information constraining the DM contribution of the host, this is a valid avenue to investigate.

Unfortunately, the maps presented in this study are unlikely to \red{provide significant constraints}; the majority of these FRBs are not localised to high enough precision, and the beam sizes likely preclude the measured column densities from being accurately representative of the line of sight to the FRB. As such, we do not attempt a correlation between excess DM and our derived \hi\ properties, leaving this for future work with a larger, well-localised sample.

% However, such a correlation is particularly ineffective given the large relative uncertainty in the Milky Way's contribution and the scatter on the Macquart relation for low-z FRBs. Of course, the FRBs which are most useful for cosmological modelling are those at higher redshifts anyway; even the best current generation observatories are unable to observe \hi\ disks with a resolution greater than 5 beams much beyond z=0.1. Though, with a larger sample it would be worth investigating to draw inferences at low 

% However, moving forward, both the localisation and sensitivity issues may be resolved, at least to an extent to make this avenue worth exploring. The deployment of the SKA, combined with the increase in sub-arcsecond localisations from upgraded FRB hunting observatories, should allow for the detailed analysis of \hi\ column densities along the lines of sight to a significant number of FRBs out to redshifts around 0.5 \citep{ska}. Such observations could therefore provide an extra layer of information to help disentangle the DM budget.

\subsection{3D Asymmetry Quantification}

Having utilised \texttt{3DACs} to measure 3D asymmetry in these host galaxies, we can assess the considerations that are likely to be important for future studies which do the same. Of course, we reiterate that only three of these targets satisfy the spatial resolution requirements outlined in \citet{Deg2020} for robust asymmetry modelling. However, using just these we can clearly show how difficult it can be to make physically meaningful asymmetry measurements. 

\begin{figure*}[t]
    \centering
    \begin{subfigure}[b]{\textwidth}
        \centering
        \includegraphics[width=0.9\textwidth]{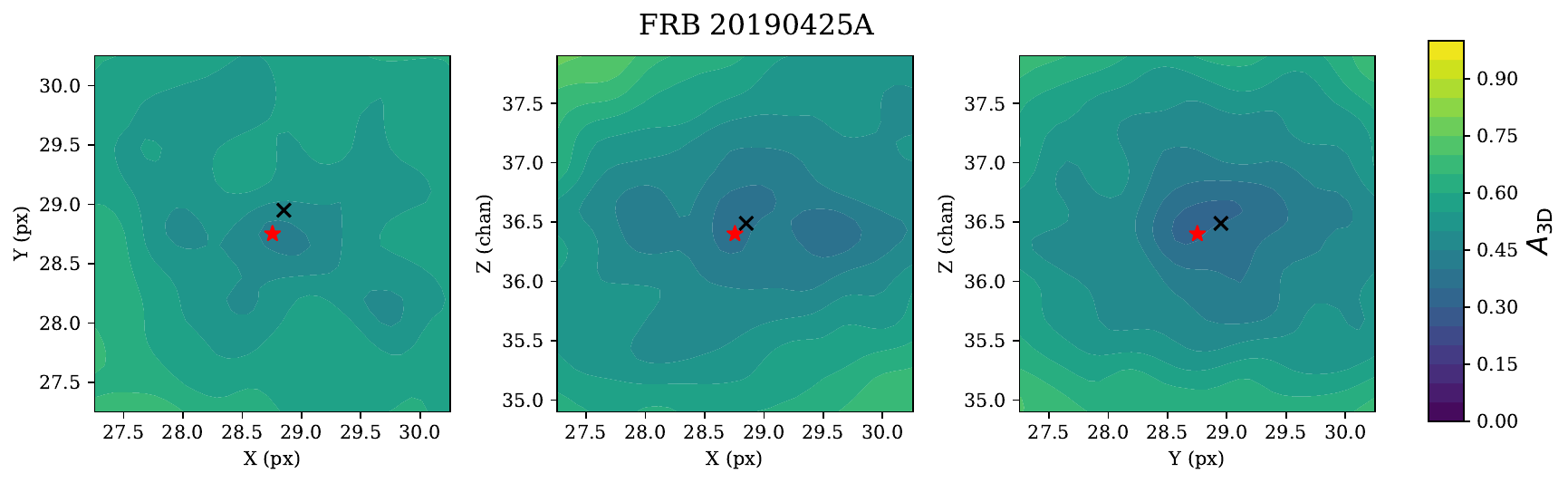}
    \end{subfigure}
    
    % \vspace{0.5cm}  % optional space between figures
    
    \begin{subfigure}[b]{\textwidth}
        \centering
        \includegraphics[width=0.9\textwidth]{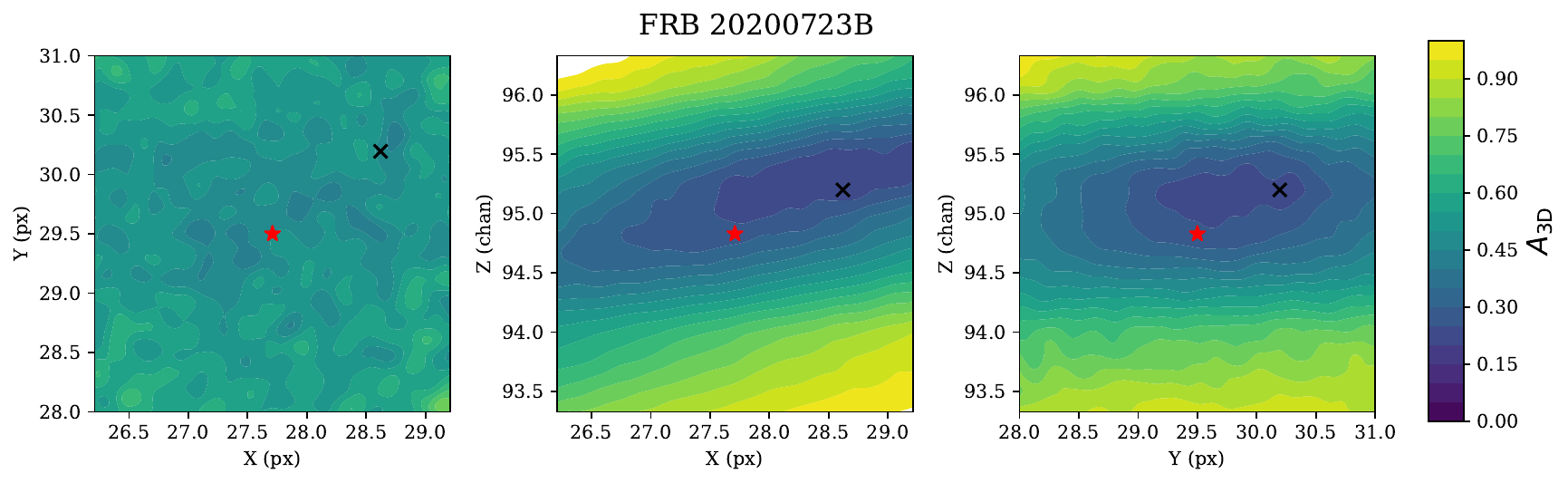}
    \end{subfigure}
    
    \caption{Smoothed sampling of the \AtD\ volumes for the hosts of FRB 20190425A and FRB 20200723B, which both pass the resolution requirements for robust \AtD\ measurement according to \citet{Deg2023}. The red star in the centres show the location of the fractional 3D voxel used for our measurements, and the black crosses designate the location of the flux-weighted centre. The slightly disturbed host of FRB 20200723B shows a much greater discrepancy between the two points, which corresponds to a significantly different measured value of \AtD.}
    \label{fig:A3D}
\end{figure*}

Figure~\ref{fig:A3D} presents a smoothed sampling of the \AtD\ space with respect to the choice of central (x,y,z) voxel, for the hosts of FRB 20190425A and FRB 20200723B. The red stars represent the chosen rotation centre (i.e.\ the flux-weighted spectral centre of the optical centre of brightness) whereas the black crosses represent simply the flux-weighted centre of the masked cube. It can be seen that very slight changes in the choice of rotation voxel can lead to significant variations in the returned \AtD\ value. 

The host of FRB 20190425A is undisturbed, and thus the discrepancy between the \AtD\ value measured using our chosen centre and the flux-weighted centre is just 0.009. However, for the slightly disturbed host of FRB 20200723B, using the latter cuts the measured value in half, from 0.210 to 0.108. This makes intuitive physical sense; a flux-weighted centre will always be biased by extended/disrupted emission. However, it is interesting to observe how significantly this impacts the asymmetry quantification even for such a low level disruption, essentially completely concealing its existence. Thus, we recommend that if kinematic modelling is not possible/feasible in order to ascertain a kinematic centre, it is certainly worth deriving the rotation voxel from optical data rather than \hi\ data.

\section{Conclusions}\label{sec:conclusion}

In this study, we have probed the \hi\ emission from \red{17} FRB host galaxies to investigate the emerging trend of disturbance and asymmetry in the population. We detect emission from \red{16} of these; the other is known to be transitioning to a quiescent stage and thus is less likely to contain a significant gas reservoir. By examining the spectra and intensity maps, and using various asymmetry quantification metrics, we identify six hosts which are disturbed, four which are not, and three which require deeper observations to draw clear conclusions on. Incorporating past results leads to a disturbed-to-settled \hi\ profile ratio for FRB hosts of \red{11:6}, which tells a much different story to the previous ratio of 5:0 \citep[or 5:1 when including the inconclusive result of ][]{Glowacki_wallaby}. \red{The ratio we find yields a p-value of 0.222 under the assumption of an underlying 1:1 distribution as is roughly expected of the background galaxy population. This clearly signifies that the observed excess of disturbance in FRB hosts is not statistically significant, dispelling any consistent underlying link between FRB progenitor formation and host galaxy merger activity. However, given 21 of 22 FRB hosts now observed in \hi\ do indeed contain emitting \hi\ reservoirs, the connection between these progenitors and star formation is further solidified.}

\red{In accordance with their optically derived stellar masses}, FRB host galaxies are mostly massive in \hi, but exhibit a spread in gas and baryon fraction that is unremarkable with respect to the background population. Hosts of repeating FRBs may be slightly less massive on average than those of apparently non-repeating FRBs, \red{though our sample size is too small to draw firm conclusions. Similar observations have been noted in recent stellar-mass studies, but significance tests in those works fail to confirm a difference; we perform comparable tests on our independent sample of repeater hosts and likewise find no statistically significant difference.} 

With a number of low-z FRBs localised since the start of this study, and with CHIME DR2 on the horizon, a significantly larger sample of local Universe FRB hosts should be observable in \hi\ in the near future. \red{Highly resolved \hi\ maps of these hosts may offer clues to aid in disentangling their contributions from the measured DM of FRBs. Future works with a large sample of arcsecond localised nearby FRBs should aim to investigate the relation between \hi\ column density and excess DM, and should also focus on probing the potential mass discrepancy between the hosts of repeating and non-repeating FRBs.}

%\endnote in some journals will behave like \footnote; and \printendnotes will not output anything. 

\section*{Acknowledgements}
We thank Ron Ekers and Kristine Spekkens for useful discussions, and Lucia Marchetti for providing access to a dataset analysed in this work. H.R. is supported by an Australian Government Research Training Program (RTP) Scholarship. M.G. and C.W.J. acknowledge support by the Australian Government through the Australian Research Council Discovery Projects funding scheme (project DP210102103). M.G. also acknowledges support through UK STFC Grant ST/Y001117/1. For the purpose of open access, the author has applied a Creative Commons Attribution (CC BY) licence to any Author Accepted Manuscript version arising from this submission. M.C. acknowledges support from the Australian Research Council Discovery Early Career Research Award (project number DE220100819). A.C.G. and the Fong Group at Northwestern acknowledges support by the National Science Foundation under grant Nos. AST-1909358, AST-2308182 and CAREER grant No. AST-2047919. 
J.X.P., A.C.G. acknowledge support from NSF grants AST-1911140, AST-1910471 and AST-2206490 as members of the Fast and Fortunate for FRB Follow-up team. 
ATD acknowledges support through Australian Research Council Discovery Project DP22010230.

This scientific work uses data obtained from Inyarrimanha Ilgari Bundara, the CSIRO Murchison Radio-astronomy Observatory. We acknowledge the Wajarri Yamaji People as the Traditional Owners and native title holders of the Observatory site. CSIRO's ASKAP radio telescope is part of the Australia Telescope National Facility (https://ror.org/05qajvd42). Operation of ASKAP is funded by the Australian Government with support from the National Collaborative Research Infrastructure Strategy. ASKAP uses the resources of the Pawsey Supercomputing Research Centre. Establishment of ASKAP, Inyarrimanha Ilgari Bundara, the CSIRO Murchison Radio-astronomy Observatory, and the Pawsey Supercomputing Research Centre are initiatives of the Australian Government, with support from the Government of Western Australia and the Science and Industry Endowment Fund. We also thank the MRO site staff. The MeerKAT telescope is operated by the South African Radio Astronomy Observatory, which is a facility of the National Research Foundation, an agency of the Department of Science, Technology and Innovation.

We acknowledge the use of the ilifu cloud computing facility—www.ilifu.ac.za, a partnership between the University of Cape Town, the University of the Western Cape, Stellenbosch University, Sol Plaatje University, the Cape Peninsula University of Technology, and the South African Radio Astronomy Observatory. The ilifu facility is supported by contributions from the Inter- University Institute for Data Intensive Astronomy (IDIA—a partnership between the University of Cape Town, the University of Pretoria, and the University of the Western Cape), the Computational Biology division at UCT, and the Data Intensive Research Initiative of South Africa (DIRISA). This work was carried out using the data processing pipelines developed at the Inter-University Institute for Data Intensive Astronomy (IDIA) and available at https://idia-pipelines.github.io. IDIA is a partnership of the University of Cape Town, the University of Pretoria, and the University of the Western Cape. This work made use of the CARTA (Cube Analysis and Rendering Tool for Astronomy) software (https://cartavis.github.io; Comrie et al. 2021), and the iDaVIE (immersive Data Visualisation Interactive Explorer) software \citep[https://idavie.readthedocs.io/en/latest/;][]{iDaVIE}  This research has made use of the NASA/IPAC Extragalactic Database (NED), which is operated by the Jet Propulsion Laboratory, California Institute of Technology, under contract with the National Aeronautics and Space Administration.

% \end{acknowledgement}

\printbibliography
\clearpage
\appendix

\section*{Appendix A: Additional Information}

\begin{table}[!ht]
\caption{\red{Extra information for Figs. \ref{fig:spectra} and \ref{fig:momentmaps}. Columns 3 and 4 display the resolution and RMS sensitivity of the bold (i.e. binned) spectra in each case in Fig.~\ref{fig:spectra}. Columns 5 and 6 display the \hi\ column / surface densities as presented in Fig.~\ref{fig:momentmaps}, starting with the $3\sigma$ level as returned by SoFiA-2, and increasing by the multiples given in column 7.}}

\label{tab:columndensity}
\begin{threeparttable}
\begin{tabular}{lc|cc|ccc}
\toprule
\headrow  & & \multicolumn{2}{c|}{Spectra} & \multicolumn{3}{c}{Moment Maps} \\ 
\cmidrule{3-7}
\headrow FRB Name & Observation & Resolution & RMS & $3\sigma$ N\textsubscript{\hi} & $3\sigma$ $\Sigma_{\hi}$ & Multiples\\
\headrow & & (\kms) & (mJy) & (\tenpow{20} cm$^{-2}$) & (\msun pc$^{-2}$) & \\ 
\midrule

FRB 20181220A & GMRT & 29.2 & 0.62 & 1.99 & 1.60 & 1,2,4,8 \\
& FAST & 12.9 & 0.30 & - & - & - \\
\midrule

FRB 20181223C & GMRT & 11.6 & 1.32 & 1.36 & 1.09 & 1,2,4 \\
& FAST & 9.7 & 0.14 & - & - & - \\
\midrule

FRB 20190418A & GMRT & 46.7 & 0.46 &  3.95 & 3.17 & 1,2,3,4 \\
\midrule

FRB 20190425A & GMRT & 14.5 & 2.34 & 1.12 & 0.90 & 1,2,4,8 \\
& FAST & 12.9 & 0.99 & - & - & - \\
\midrule

FRB 20200223B & GMRT & 10.4 & 0.50 & - & - & - \\
& FAST & 12.9 & 0.36 & - & - & - \\
\midrule

FRB 20200723B & MeerKAT & 44.1 & 3.26 & 1.26 & 1.01 & 1,2,4,8 \\
\midrule

FRB 20201123A & MeerKAT & 132.3 & 0.14 & 2.13 & 1.71 & 1,2,4 \\
\midrule

FRB 20210405I & MeerKAT & 11.0  & 0.93 & 0.67 & 0.53 & 1,2,4,8 \\
\midrule

FRB 20211127I & MeerKAT & 11.0 & 0.81 & 0.93 & 0.74 & 1,2,4,8,16 \\
\midrule

FRB 20211212A & MeerKAT & 22.0 & 0.35 & 0.28 & 0.23 & 1,2,3 \\
\midrule

FRB 20231229A & MeerKAT & 21.3 & 1.65 & 0.27 & 0.22 & 1,2,4 \\
\midrule

FRB 20231230A & MeerKAT & 44.1 & 1.42 & 0.31 & 0.25 & 1,1.3,1.6,1.9 \\
\midrule

FRB 20240201A & MeerKAT & 22.0 & 0.35 & 1.24 & 0.99 & 1,2,4,8,16 \\
& FAST & 12.9 & 0.20 & - & - & - \\
\midrule

FRB 20240210A & MeerKAT & 11.0 & 0.95 & 2.39 & 1.91 & 1,2,4,8 \\
\midrule

FRB 20240312D & MeerKAT & 22.0 & 0.34 & 1.20 & 0.96 & 1,2,4,8 \\
\midrule

FRB 20240615B & MeerKAT & 11.0 & 0.43 & 0.27 & 0.22 & 1,2,4 \\
\midrule

FRB 20240615B & MeerKAT & 10.4 & 5.94 & 0.27 & 0.22 & 1,2,4 \\
\midrule

\bottomrule
\end{tabular}
\end{threeparttable}
\end{table}

\clearpage
\clearpage

\section*{Appendix B: Asymmetry Metrics vs. Resolution}\label{appendixB}

\begin{figure}[h!]
    \centering
    \includegraphics[width=0.97\linewidth]{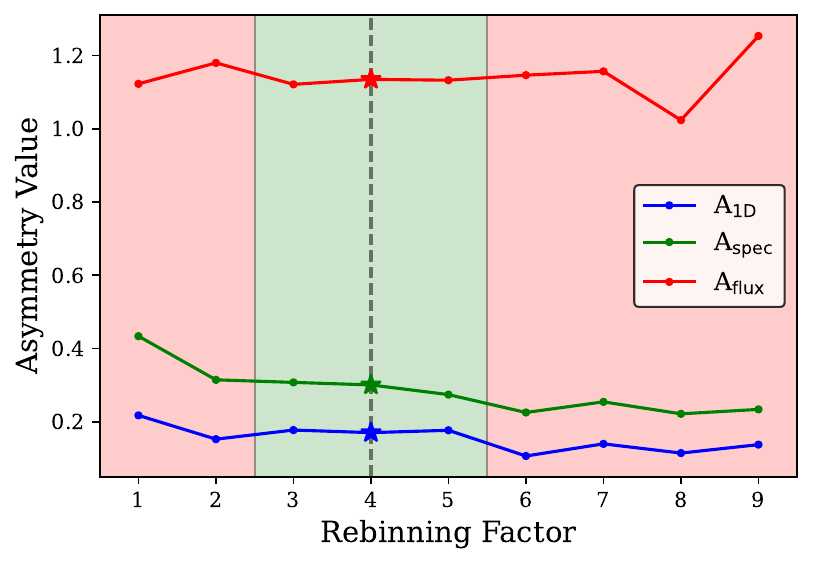}
    \caption{1D asymmetry metrics for FRB 20210405I at various resolutions.}
    \label{fig:asymdiagnostic}
\end{figure}

As outlined in Section \ref{subsec:asym_metrics}, measuring the asymmetry metrics at various resolutions is particularly important for lower SNR targets. Fig.~\ref{fig:asymdiagnostic} demonstrates our methods for determining the optimal resolution at which to take measurements. Here we take a high resolution, lower SNR target (the host of FRB 20210405I), and display the values of \Aflux, \Aspec, and \AoD\ as a function of rebinning factor. The values appear to be particularly unstable at low rebinning factors (due to noise) and high rebinning factors (due to resolution); these regimes are shown in red. The green region appears the most stable in all three metrics, with the central rebinning factor of 4 thus chosen as the one we use in our final asymmetry measurement.

\newpage
-
\newpage

\section*{Appendix C: Kinematic Modelling of FRB 20200723B}

\begin{figure}[hbp]
    \centering
    \includegraphics[width=1.35\linewidth]{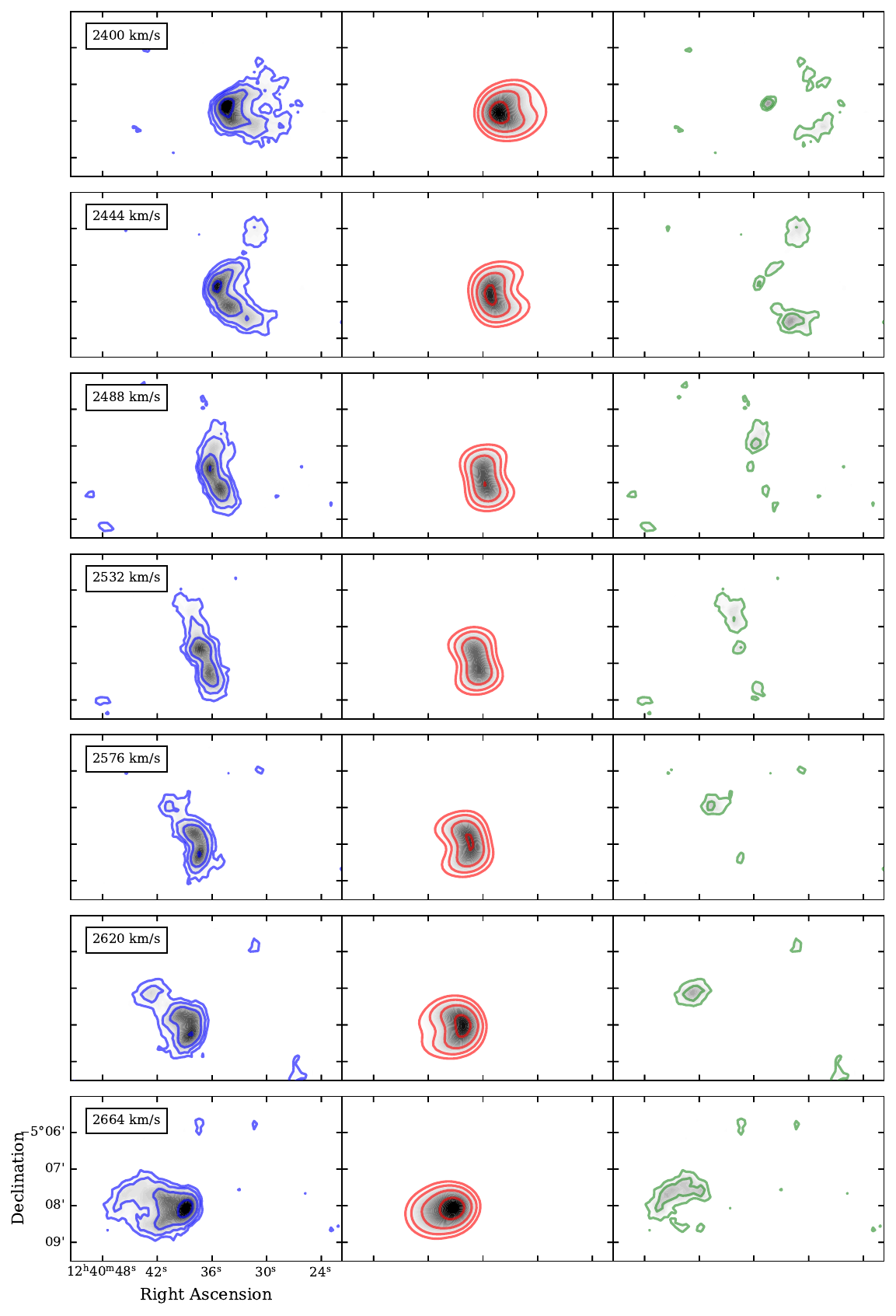}
    \caption{Comparison between observations of the host of FRB 20200723B (left) and a smooth rotating disk model from \texttt{3DBarolo} (centre), with residuals shown (right).}
    \label{fig:barolo}
\end{figure}

\newpage
-
\newpage
\section*{Appendix D: Host Disturbance vs. \hi\ Properties}

\begin{figure}[h!]
    \centering
    \includegraphics[width=0.92\linewidth]{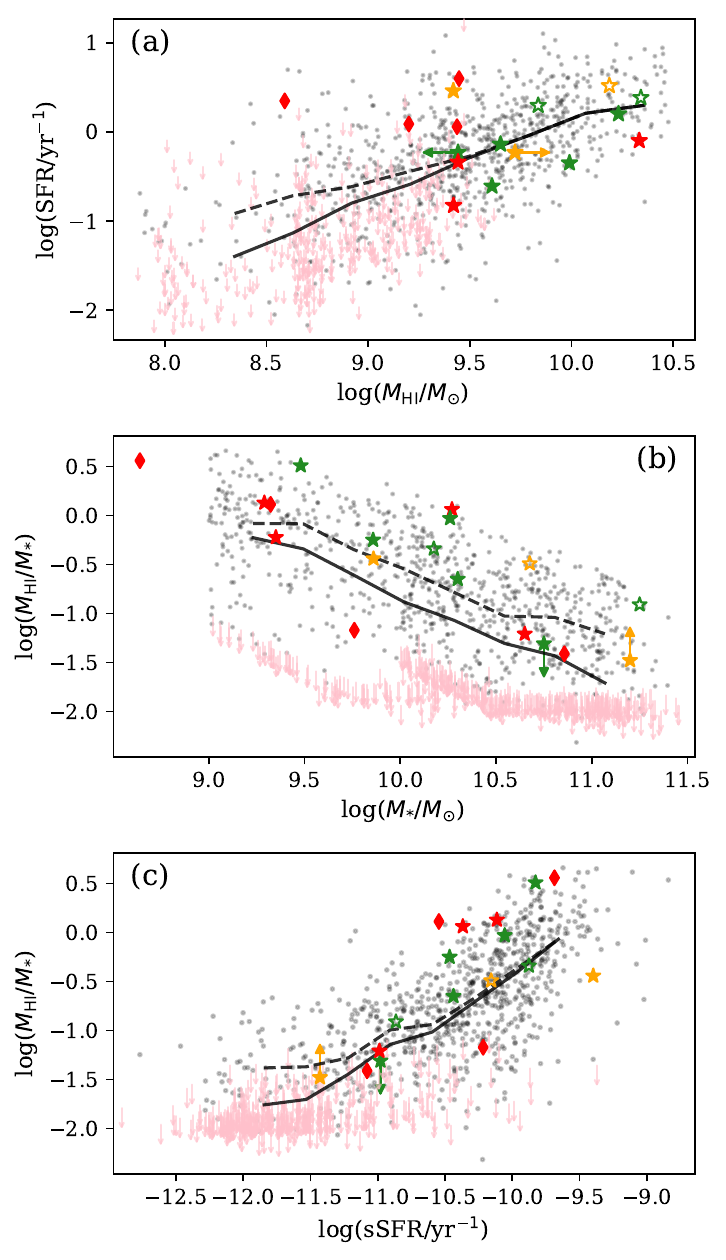}
    \caption{Same as Fig.~\ref{fig:hi_properties} but with targets coloured by disturbance classification.}
    \label{fig:asym_hi}
\end{figure}

Fig.~\ref{fig:asym_hi} presents an identical plot to Fig.~\ref{fig:hi_properties}, just with FRB hosts colour-coded by our disturbance classification. Red targets are those we consider disturbed, orange are potentially disturbed, and green are settled. 
 
\end{document}